\documentclass{template}
\usepackage{xcolor}
\usepackage{float}
\usepackage{hyperref}
\usepackage{amsfonts}
\hypersetup{
  colorlinks   = true, 
  urlcolor     = blue, 
  linkcolor    = blue, 
  citecolor    = blue   
}
\begin{document}

\pagestyle{fancy}
\rhead{\includegraphics[width=2.5cm]{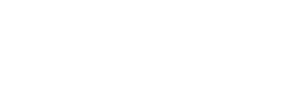}}

\title{Meta-analysis of literature data in metal additive manufacturing: What can we (and the machine) learn from reported data?}

\maketitle


\author{Raymond Wong}\textsuperscript{a,$\ast$},
\author{Anh Tran}\textsuperscript{b},
\author{Bogdan Dovgyy}\textsuperscript{c},
\author{Claudia Santos Maldonado}\textsuperscript{a},
\author{Minh-Son Pham}\textsuperscript{a,$\ast$}



\begin{affiliations}

\textsuperscript{a}
Department of Materials,
Imperial College London,
London, SW7 2AZ, UK 

\textsuperscript{b}
Scientific Machine Learning, 
Sandia National Laboratories,
Albuquerque,
NM 87123, USA 

\textsuperscript{c}
International Additive Manufacturing Group, Domaniewska 3,
05-800 Pruszkow,
Poland 

\textsuperscript{$\ast$} Corresponding authors. \\
Email Addresses: r.wong21@imperial.ac.uk and son.pham@imperial.ac.uk

\end{affiliations}


\keywords{Data analytics, Machine learning, Processing maps, Additive manufacturing, Alloys}

\begin{abstract}

Obtaining in-depth understanding of the relationships between the additive manufacturing process, microstructure and mechanical properties is crucial to overcome barriers in additive manufacturing (AM). Over the past decades, there have been significant studies in AM, providing considerable amount of data available for examination of such relationships broadly across many literature studies. In this study, database of metal AM was created thanks to a large number of literature studies. Subsequently meta-analyses on the data was undertaken to provide insights into whether such relationships are well reflected in the literature data. The analyses help reveal the bias and what the data tells us, and to what extent machine learning can learn from the data. The first major bias is associated with common practices in identifying the process based on optimizing the consolidation. Most data reports were for consolidation while data on microstructure and mechanical properties was significantly less. In addition, only high consolidation was only provided. Machine learning trained on the data was not able to learn the process - consolidation relationship in the medium and low end ranges of consolidation. The common identification of process maps based on only consolidation also poses another bias because mechanical properties that ultimately govern the quality of an AM build are controlled not only by the consolidation, but also microstructure. However, the number of studies on quantifying the microstructure was extremely low, limiting the learning of the microstructure - mechanical properties relationships. Meta-analysis of the literature data also shows weak correlation between input (i.e. process parameters) with output (i.e. consolidation and mechanical properties). This weak correlation is attributed to the stated biases and the highly non-monotonic and non-linear relationships between the process and quality variables. Fortunately, machine learning models trained on the data capture well the interactions between process parameters are influential in the output, and predicts accurately the yield stress, suggesting that the correlation between process, microstructure and yield strength is well reflected in the data.  Last but not least, due to the current limitation in the process map identification, we propose to identify the process map on the basis of not only the consolidation, but also mechanical properties. Such a identification shows that 316L and Inconel have much larger process map (i.e. highly printable) in comparison to the Ti6Al4V, Hastelloy-X and Inconel 625. 

\end{abstract}


\section{Introduction}

Additive manufacturing (AM) has a potential to revolutionize the manufacturing industry by offering an efficient and cost-effective method for fabricating complex structures, providing advantages over other manufacturing techniques \cite{Debroy2018}.
Despite its advantages, metal AM presents major challenges due to the formation of defects and undesirable microstructure, affecting the mechanical performance and reliability of final products in applications \cite{Lewandowski2016,Gu2012}. In particular, extreme interactions between the energy beam and materials and associated complex thermal conditions in AM cause difficulties in understanding of the underlying relationship between alloy composition, microstructure and properties of additively manufactured alloys \cite{Schmelzle2015ReDesigningManufacturing, King2015LaserChallenges, Zhang2019MetalChallenges, Maleki2021SurfaceOpportunities}. Overcoming such challenges requires fundamental understanding of the process-microstructure-property relationship that will assist the development of (1) forward engineering (predicting the mechanical properties of a given alloy for a specific set of process parameters), and (2) reverse engineering (identifying the alloy composition and corresponding process parameters for a given set of properties).

Over the past decade, there has been a considerable number of studies reporting the relationships between process, microstructure and properties (PMP). Meta-analysing a considerable large data reported by many research groups has potential to unravel such important PMP relationships and identify the good and not good practices in studying metal additive manufacturing. Such meta-analysis will also allow us to identify the bias and associated implications in our learning of the PMP relationships. Meta-analysis has becomes powerful in providing invaluable insights that may not be immediately apparent thanks to advances in data analytics (DA) and machine learning (ML)  \cite{Segaran2009BeautifulData,RunklerDataEdition,Rasmussen2019}.

Despite the significant use of DA (and ML) in many fields, such as healthcare and online shopping, the use of data analytics for AM is still in its early stages \cite{Horvitz2015DataGood}. Although some successes is shown in these data-driven studies (for example, mechanical properties based on simulated microstructure or process parameters \cite{Garg2014AProcess,Zhang2019DeepModeling, Bayraktar2017ExperimentalNetworks,Garg2014State-of-the-artProcesses}, melt pool dimensions for a given composition or process parameters \cite{Lee2019DataManufacturing,Kamath2016DataMelting, Yang2020FromMethod,Tapia2018GaussianSteel, Yang2021IDETC/CIEMANUFACTURING, Meng2020ProcessModel,Wang2020UncertaintyManufacturing}, and porosity based on process parameters used or images of the build \cite{Rankouhi2021CompositionalParameters,Imani2019DeepControl,Liu2020Machine-learningMechanisms,Garg2015MeasurementMethods, Shin2021OptimizingIntelligence,Tapia2016PredictionModels,WangYNet:Evolvement}, such studies were only based on limited sets of conditions. While comprehensive material databases such as Materials Project, AFLOWLIB and OQMD are available for functional material properties, there has been little effort to develop similar database targeted for AM \cite{Taylor2014AConsortium,Jain2013Commentary,Saal2013MaterialsOQMD}. Notably, NIST and the now discontinued Citrination have placed efforts to address the lack of a comprehensive database for AM. Nevertheless, it is still inadequate to be used for big data analysis at the scale of other fields \cite{nist,citrination}.
Fortunately, there are a considerable amount of data reported in literature studies for AM in the past decades \cite{nist,Wang2022,Popova2017,Wang2020,Razvi2019,Zhang2019,Lyu2022}. Such literature studies contain valuable data reflecting some key aspects of the PMP relationships in AM alloys. However, there had not been collection and structuring of rich data available in literature. This study firstly creates considerable database by collecting, subsequently structuring and organising data available in literature with focus on the laser powder bed fusion (LPBF) that is currently the most used AM method in fabricating metallic alloys. Secondly, the study carries out in-depth examination of the data to identify biases in the reported data that may hinder the learning of the PMP relationship for AM alloys. This will be achieved by conducting an analysis of the correlation, principal component and sensitivity analyses. Furthermore, ML algorithms will be used to test the current performance of the obtained dataset and provide bases to analyse the sensitivity of process parameters on the predicted consolidation and mechanical properties of the trained ML models. Last but not least, the study will identify process window maps that are optimized on the basis of not only the consolidation, but also key mechanical properties that are crucial for structural applications.
Consequently, the use of DA and ML on a significantly large number of studies can provide invaluable insights relating to the PMP relationships in particular processing parameters and their effects on the printed product, assisting the AM users to optimize processing parameters. Such knowledge will also assist the AM users in assessing the printability of existing alloys and accelerating the search for new printable alloys with desired properties  \cite{Dovgyy2021Alloy,Tang2021Alloys,Thapliyal2020AnManufacturing,Mycroft2020AProcesses,Yan2018Data-drivenManufacturing,Popova2017Process-StructureData}.

\section{Methods}
\label{sec:ch3_sec1}

An extensive literature search has been undertaken to create significant datasets from published data reported in literature.
Over \textit{2000} data entries has been obtained for commonly printed alloys such as Ti6Al4V, Inconel 718, Inconel 625, Hastelloy X and 316L usig the laser powder bed fusion. The data collection from literature reports included processing parameters, consolidation and mechanical properties of as-built conditions and post processing information if available (see \textbf{Figure \ref{fig:Overall Histogram}}). 
Powder bed fusion (PBF) is currently the most commonly used AM technique and has most comprehensive published data in comparison to other variations such as DED. Therefore, the data collection was done for PBF with focus on LPBF. The knowledge gained from this analysis will be applicable to electron beam powder bed fusion.

The work focuses on exploratory data analysis (EDA) to establish relations between different variables in the dataset to understand the characteristics (including biases and limitations of current data reporting practices) and underlying process - consolidation - mechanical properties relationships hidden in the obtained datasets. This includes analysis of the reported data and uncovering relations between input parameters and output properties (including consolidation, yield stress (YS), ultimate tensile stress (UTS) and elongation). Following this, analysis of optimized process windows has been undertaken to gain insights into the relation between process parameters, material and properties of the selected alloys. 
Finally, trained supervised ML models have been used to investigate the impact of current limits of the reported data (including common practices) in the learning by machine.

\subsection{Explanatory Data Analysis} 
\label{sec:ch3_subsec11}

Correlation techniques have been conducted to investigate the underlying correlation, issues and characteristics within the dataset in order to uncover deeper relationships between data features. The most commonly used techniques are Pearson's correlation coefficient and Spearman's correlation coefficient. Pearson's is used to capture linearity correlation between two variables and is defined as


\begin{equation}
\label{eq:Pearsons}
     r_p = \frac{\sum(x - \overline{x})(y - \overline{y})}
     {\sqrt{\sum(x - \overline{x})^{2} \sum(y - \overline{y})^{2} }} \bigskip
\end{equation}


where $x$ and $y$ are the individual values of the two variables,
      $\overline{x}$ and $\overline{y}$ denote the mean of the two variables in the dataset \cite{DouglasC.Montgomery2012IntroductionAnalysis,Hogg2018IntroductionStatistics}. 
Spearman's correlation is used to capture the strength of a monotonic relation between two variables and is given by


\begin{equation}
\label{eq:Spearmans}
     r_s = 1 - 6 \sum \frac{d^{2}}
     {N(N^{2} - 1)} \bigskip
\end{equation}


where $d$ denotes the difference between the two variables ranks, and $N$ denotes the sample size \cite{DouglasC.Montgomery2012IntroductionAnalysis,Hogg2018IntroductionStatistics}. The two correlations are typically used in conjunction with each other to capture both types of correlation which the other would not be able to capture.

The established relationship between the variation of processing parameters and the resulting defects suggests that the processing parameters are subject to multicollinearity. An example of the correlation has been demonstrated by Gordon et al. when optimizing laser power and speed, correlation is shown due to the underlying dependencies of the objective when optimizing print quality by balancing the two parameters \cite{Gordon2020DefectManufacturing}.
Multicollinearity among the processing parameters (predictors) may have a negative impact on the performance of ML models, as changes in one process parameter will inherently influence the values of one or more other process parameters. Hence, individual changes in process parameters and its effect on the dependent variables will be difficult to differentiate between the effect of other process parameters \cite{Chan2022}.
To address this, variance inflation factor (VIF) has been used to quantify the degree of multicollinearity. VIF measures the degree of collinearity of each predictor by forming a regression of one predictor to all the other predictors, calculated as follows


\begin{equation}
\label{eq:VIF}
     VIF_n = \frac{1}{1 - R_n^2} \bigskip
\end{equation}


where $R_n^2$ is the coefficient of determination of the auxiliary regression for the $n$th predictor. A VIF of 1 indicates no collinearity; whereas, VIF exceeding 5 or 10 indicates high levels of multicollinearity as a standard practice.

\subsection{Process Window Identification} 
\label{sec:ch3_subsec13}

To better understand the influence of multiple process parameters on the print quality, a window (i.e. heat map) of process parameters optimized for a separate output (or a combined output that consolidates all considered separate output variables) have been generated. This approach analyzes the raw data without the influence of ML and can provide users with predictions of output qualities purely based on reported data. Because laser power and laser speed are the most varied input parameters in literature (\textbf{Figure \ref{fig:Input Histogram}}), heatmap was constructed on the basis of these two separate input variables.
Two types of heatmaps with have been generated, $Hmap_1$ and $Hmap_2$. $Hmap_1$ is based on an individual quality (i.e. output), either YS, average work-hardening, elongation or consolidation ($z$-axis) for a given laser speed ($x$-axis) and laser power ($y$-axis). Such that four $Hmap_1$ will be generated for each output variable.
As discussed earlier that consolidation (or any single quality variable) is not sufficient to reflect the quality. Therefore, $Hmap_2$ is constructed to identify a process window that can achieve a combined variable that includes all the considered individual four output variables.

\begin{figure*}[h!]
\centering
\includegraphics[width = 0.5\hsize]{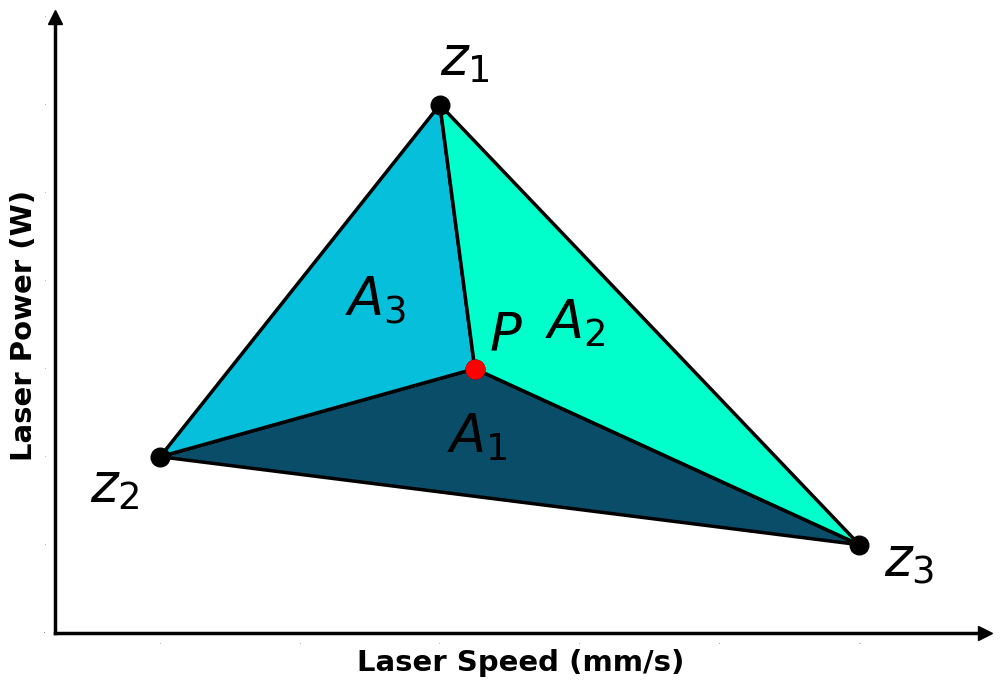}
\caption{Illustration of Barycentric interpolation to obtain value at point $P$, given three known $Z$ values, corresponding to YS, average work-hardening, elongation or consolidation.}
\label{fig:Barycentric}
\end{figure*}

The generation of $Hmap_1$ consists of interpolating the data points within its convex hull to obtain $z$ values at the point of interest $P$ within this region. As the data points are not uniformly distributed, Barycentric interpolation was chosen over other methods as it accounts for the distances between neighboring data points, leading to more accurate and smoother estimates of the function value. In particular, the method employs a series of interpolations over small regions of parameterized triangles formed between data points \cite{Virtanen2020SciPyPython,Barber1996TheHulls,Berrut2004BarycentricInterpolation}. The interpolation of $P$ uses Barycentric coordinates ($\lambda$) of the parameterized triangles, given by


\begin{equation}
\label{eq:Barycentric Lamda}
     \lambda_n = \frac{A_n}{A_{total}} \bigskip
\end{equation}


such that $\lambda$ represents the proportional size of the interior triangles formed by connecting point $P$ with each vertex of the parameterized triangle. Where $A_n$ denotes the area of an interior triangle and $A_{total}$ represents the total area of the parameterized triangles. An illustration of the Barycentric coordinates is shown in \textbf{Figure \ref{fig:Barycentric}}.
Using these coordinates, $P$ is determined as a weighted average dependent on the distance of the neighboring $z$ values as follows


\begin{equation}
\label{eq:Barycentric}
     P = \sum_{n=1}^{3} \lambda_n z_n \bigskip
\end{equation}


To generate $Hmap_2$, all interpolated values for $Hmap_1$ of the same material has been normalized with a min-max scaler and summed with equal weighting to create a map which shows the optimized process parameter regions. Thus, the map considers process - consolidation - mechanical properties relationships as opposed to single properties like consolidation. The min-max scaler ensures all values for the map ranges from 0 to 1, such that the scaling of output qualities will contribute equally to the summarized map \cite{Albon2018MachineLearning}. However, if any output quality is considered to be more important, a higher weighting factor can be added to allow more contribution from that individual output to the combined quality.


\subsection{Machine Learning} 
\label{sec:ch3_subsec12}

Principal component analysis (PCA) was used to reduce the dimensionality of the input data. The efficacy of the dimensionality reduction has been investigated to study the use of volumetric energy density (VED) that is commonly used metric to optimize print quality. Furthermore, multiple supervised ML techniques have been employed to examine the accuracy and effects of data biases and limitations on the ML performance. Machine learning models employed include decision tree, random forest, support vector machine, neural network, XGBoost, LightGBM and CatBoost. 
Based on the observations of non-Gaussian distributed data and sensitivity analyses shown in \textbf{Figure \ref{fig:Input Output Histograms}} and \textbf{Table \ref{tab:VIF}} respectively, the selection of these models focuses on methods which doesn't assume Gaussian distribution and is less susceptible to multicollinearity \cite{Brownlee2017,Geron2019,hastie01statisticallearning}. To employing ML models on the dataset, it is essential to pre-process the data by converting it into a machine-readable format. Steps taken includes standardizing, dropping a set of data if the entry contains any missing values and applying \textit{One Hot Encoding} the material and treatment data.
\textit{One Hot Encoding} creates a new category in a binary format for unique inputs within the material and treatment information dataset. This is necessary as the machine is incapable of taking string inputs, thereby converting the data to numerical inputs allows the machine to consider material type and treatment information. Furthermore, the binary format of the categorical values avoids asserting false linear relationships which \textit{label encoding} may induce as the ML models may misinterpret the numerical labelling as rankings \cite{Albon2018MachineLearning}.
Following this, the training, validation and test data has been split into a 80\%/10\%/10\% ratio to evaluate the final performance of the model.

About 80\% of more than 2000 data entries obtained from literature was dropped due to the incompleteness for example missing mechanical properties data. Efforts to impute the missing data has been undertaken; however, the significant amount of incomplete data resulted in poor performance. Therefore, the model has been trained on over 300 points (of full data) after preprocessing the data
To produce an effective model, all aforementioned models were hyperparameter-tuned with 500 trials and a RMSE loss function; the selection of this loss function is to reduce the effect of large outliers during the prediction and training phase by penalizing larger errors due to outliers \cite{Chai2014RootMAE}.

\subsection{Sensitivity Analysis } 
\label{sec:ch3_subsec14}

It is common to simply trust a ML model because of high accuracy predictions and omit the interpretation of what the model has learnt. However, it is difficult to interpret what the machine has learned from the data. 
Sensitivity analysis enables interpretability of ML models by revealing the influence of model inputs on the model outputs.
This analysis will allow quantification of the influence of process parameters and how this varies across different models. Subsequently, the resulting values can be used to determine whether a ML model is able to reflect the underlying science between process and properties \cite{ElBilali2023,Kapusuzoglu2021}.
Sobol sensitivity analysis is a variance-based approach to quantify how the uncertainty in individual model inputs contributes to the uncertainty of model outputs; the method offers two ways of sensitivity analysis.
The main effect sensitivity index determines the individual effects of a single process parameter input, without considering the interaction of this parameter with other process parameters. This can be obtained by


\begin{equation}
\label{eq:Sobol First}
     S_i = \frac{{\mathbb{V} [\mathbb{E}(Z|x_i)]}} {{\mathbb{V}(Z)}} \bigskip
\end{equation}


such that $S_i$ measures the variance of the conditional expectation $\mathbb{V} [\mathbb{E}(Y|x_i)]$, relative to the total variance $\mathbb{V}(Z)$ for output $Z$, given an input $x_i$ \cite{Sobol2001, Saltelli2010}.
Total effect sensitivity index ($S_{T_i}$) determines the combined effects of a single process parameter input, considering their interactions with other process parameters. This can be calculated


\begin{equation}
\label{eq:Sobol Total}
     S_{T_i} = \frac{{\mathbb{E} [\mathbb{V}(Z|x_{\sim i})]}} {{\mathbb{V}(Z)}} \bigskip
\end{equation}


where $\mathbb{V}(Z|x_{\sim i})$ denotes the variance of conditional expectation for output $Z$  for all inputs except the $i$-th element \cite{Sobol2001, Saltelli2010}.
This can also be expressed as the sum of the first ($S_i$) and higher order interactions with other model inputs. For example, for laser power, this is expressed as


\begin{equation}
\label{eq:Sobol Total}
     S_{T_P} = S_{\{P\}} + S_{\{P,v\}} + S_{\{P,h\}} + S_{\{P,t\}} + S_{\{P,v,h,t\}} \bigskip
\end{equation}


where $P$, $v$, $h$, and $t$ denotes laser power, laser speed, hatch spacing and layer thickness, respectively. Such that $S_{\{P,v\}}$ is the measure of the effect from the co-variance of both laser power and laser speed.
The calculation of $S_i$ and $S_{T_i}$ typically requires approximation by Monte-Carlo sampling. This will be achieved by generating 100,000 model inputs using Saltelli sampling scheme 
\cite{Saltelli2010}. Following this, the trained ML models will use the generated model inputs to predict the interested properties. The generated set of model inputs will provide a diverse and suitable set of process parameter combinations ensuring that the analysis encompasses a broad range of parameter values and interactions. By exploring the parameter space more comprehensively, this enhances the reliability of the calculated indices of the sensitivity analysis \cite{Saltelli2010,Tran2022,ElBilali2023,Kapusuzoglu2021}.

\section{Results and Discussion} 
\label{sec:ch3_sec2}

\subsection{Explanatory Data Analysis} 
\label{sec:ch3_subsec21}

\begin{figure*}[h!]
\centering
\includegraphics[width = 0.9\hsize]{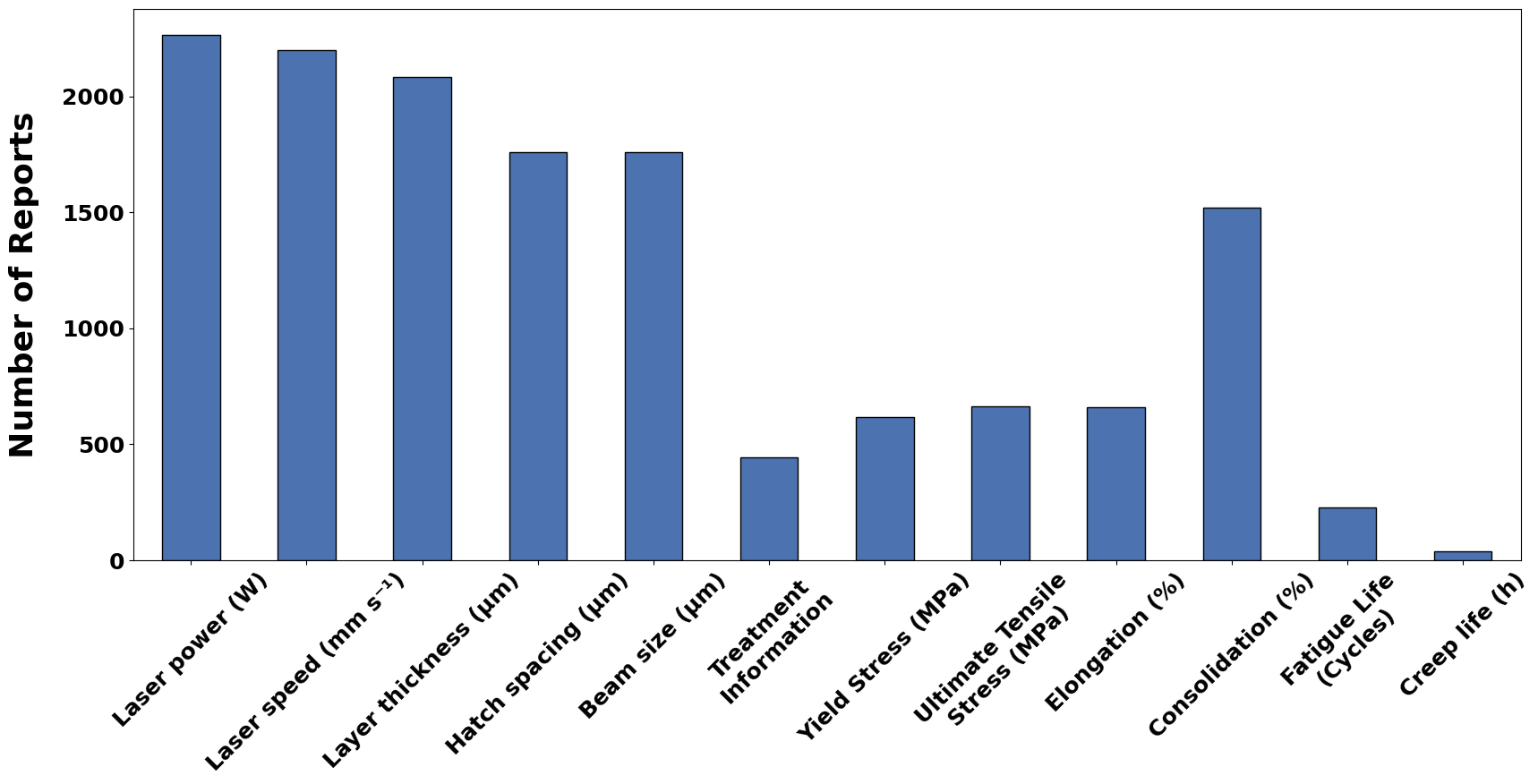}
\caption{Histogram of reported data from LPBF studies for the collected dataset.}
\label{fig:Overall Histogram}
\end{figure*}

The distribution of the compiled dataset is summarized in \textbf{Figure \ref{fig:Overall Histogram}}.
The histogram emphasizes a bias in data reporting, most of studies reported data concerning the consolidation, but much fewer on mechanical property data that are, in fact, key indicators (much more than the consolidation) of the build quality for structural applications. The histogram shows that if a machine learning algorithm was trained on the literature reports, it might understate the importance of mechanical properties, even though these qualities accounts for the print quality and is significant in practice. In addition, although there is increasing a number of studies reporting the microstructure information, most of microstructure data is qualitative and quantitative data of microstructure remains very rare. Consequently, due to insufficient data reported for the microstructure and mechanical properties, ML would not able to learn the PMP relationship and would fail to provide accurate prediction of the mechanical data for given process parameters.
Even on the consolidation data, there are limitations in method used to measure the consolidation. Hence the consolidation data contains considerable biases. Consolidation has been often measured by quantifying the density of defects such as porosity. While the optical (or electron microscopy) method can observe small defects, it is not effective in quantifying the 3D spatial distribution. By contrast, the X-ray tomography has limitations in observing fine defects. Therefore, reported data of porosity do not fully reflect the consolidation of an AM build. Also, different measurement methods can yield different results of porosity \cite{Spierings2011ComparisonParts}, resulting in inconsistent results. Moreover, even if the consolidation is accurately measured, it is known that the mechanical properties and performance of any mechanical/structural component are greatly dependent on other governing factors such as microstructure. Thus, the use of consolidation alone does not reflect well the quality of an AM build. Consequently, qualities that govern the mechanical performance such as YS, UTS, elongation, toughness, fatigue and creep that are important indicators of the load-bearing capacity and performance in structural applications should be included in the qualification of final products. Unfortunately, there have been insufficient data regarding toughness, fatigue and creep for meaningful analyses \textbf{Figure \ref{fig:Overall Histogram}}. Therefore, this study only includes YS, UTS and elongation in consideration. To improve our consideration of the build quality, we proposed to introduce an additional output variable that is the average work-hardening. Work-hardening is an important parameter reflects the energy absorption capacity of a metal according to the Considère hypothesis \cite{doitpoms}. The variable also indicates the tendency of a metal against the localisation that is one of the main mechanisms responsible for crack initiation \cite{jalal2023}. Average work-hardening has been calculated by computing the difference between UTS and YS, divided by elongation. Together with the use of the YS, the introduction of the average work-hardening makes the UTS redundant and no longer needed. 



\begin{figure}[h!]
\centering
\begin{subfigure}{0.95\linewidth}
    \caption{}
    \includegraphics[width = 0.95\hsize]{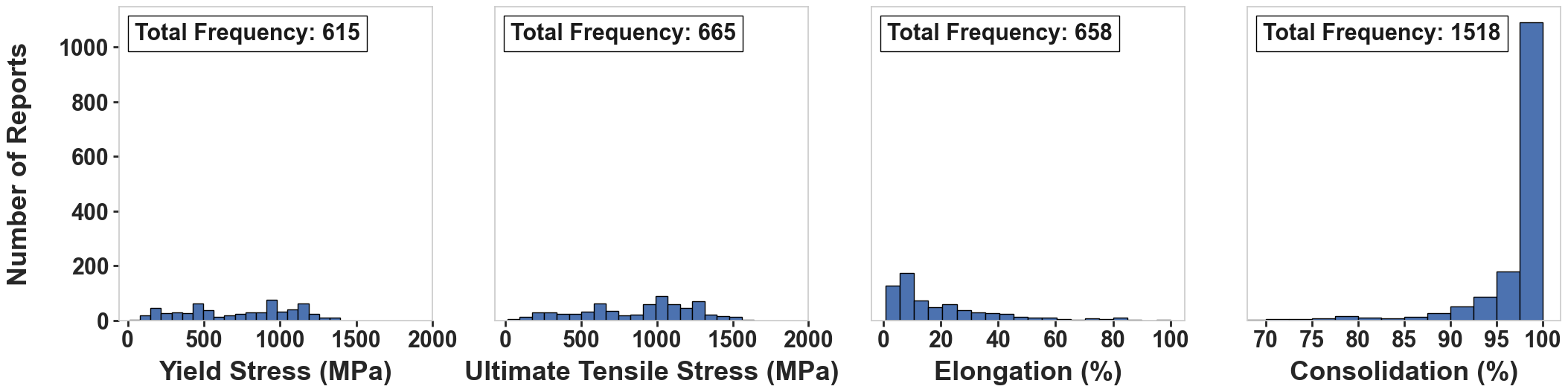}
    \label{fig:Output Histogram}
\end{subfigure}
\begin{subfigure}{0.95\linewidth}
    \caption{}
    \includegraphics[width = 0.95\hsize]{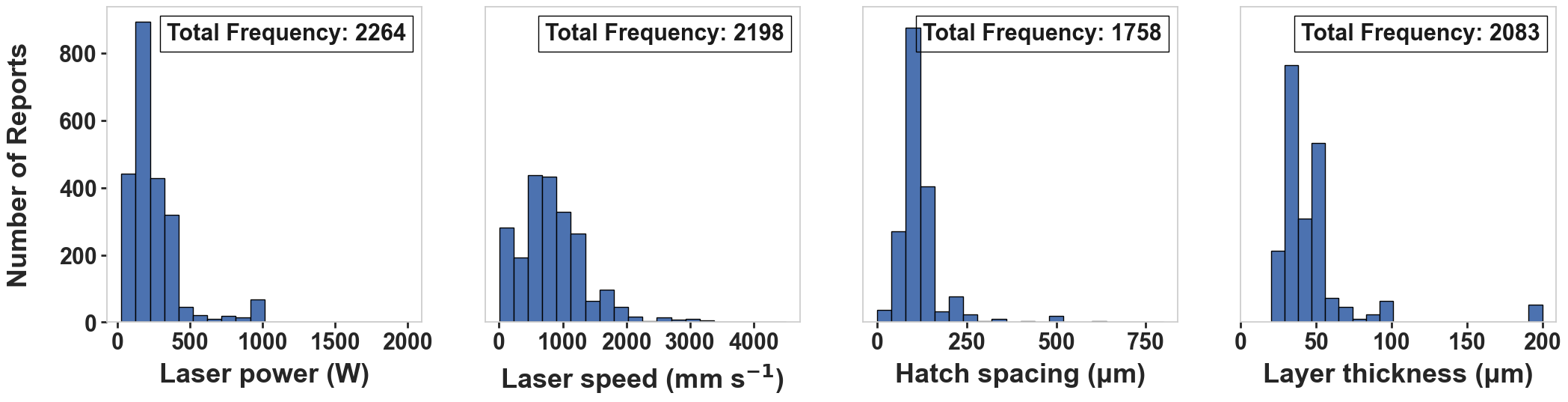}
    \label{fig:Input Histogram}
\end{subfigure}
\caption{\textbf{(a)} Histogram of reported output properties from the collected LPBF dataset. \textbf{(b)} Histogram of reported processing parameters from the collected LPBF dataset.}
\label{fig:Input Output Histograms}
\end{figure}

Analysis of the collected data showed a significant bias for high consolidation values, with over 80\% of studies only reporting results with consolidation values above 95\%. The highly skewed distribution is displayed in \textbf{Figure \ref{fig:Output Histogram}}. The reporting of process parameters for high consolidation data only (all is above 70\% with most of reports on above the 98\%) without sufficient data for low consolidation creates a major bias. It will be shown later in the correlation analyses that this bias causes the data not to reflect the relationship between the process parameters and the consolidation well. Therefore, this bias could limit the machine in learning of the full relationship between the process parameters and consolidation. Training the machine using ML models on this skewed data causes the machine not to be able to accurately predict the process parameters for low consolidation, negatively affecting the performance of ML models. It is, therefore, calling the AM community to publish the process parameters that produce low consolidation alongside with currently reporting of high consolidation values. This is an essential step in formulating an extensive dataset as a ML model will produce biased results if trained with a dataset only containing ``good'' data \cite{James2021AnEdition,RunklerDataEdition,GrusDataScratch}.


\begin{figure}[h!]
\centering
\begin{subfigure}{0.8\linewidth}
    \caption{}
    \includegraphics[width=\linewidth]{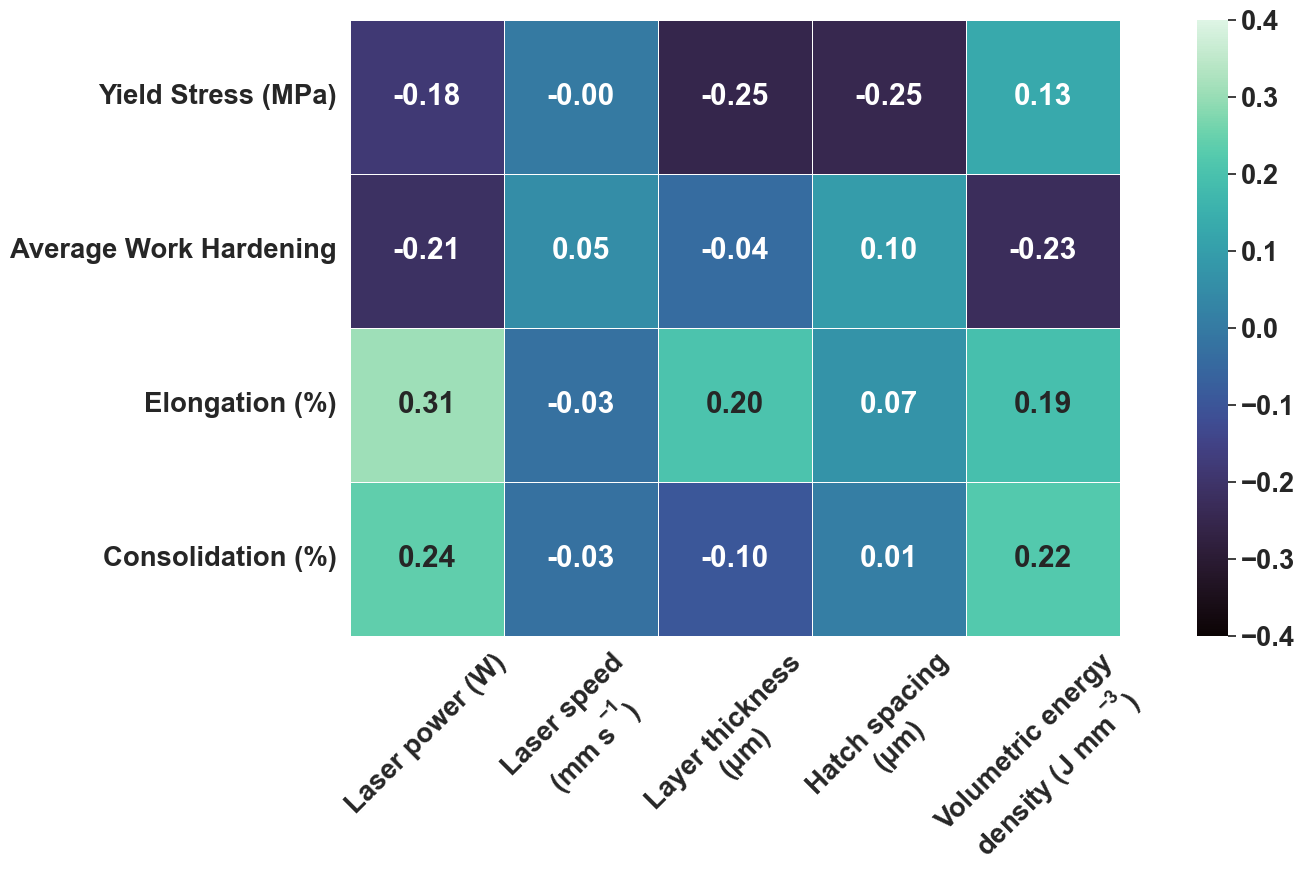}
    \label{fig:Spearmans Output Heatmap}
\end{subfigure}
\begin{subfigure}{0.8\linewidth}
    \caption{}
    \includegraphics[width=\linewidth]{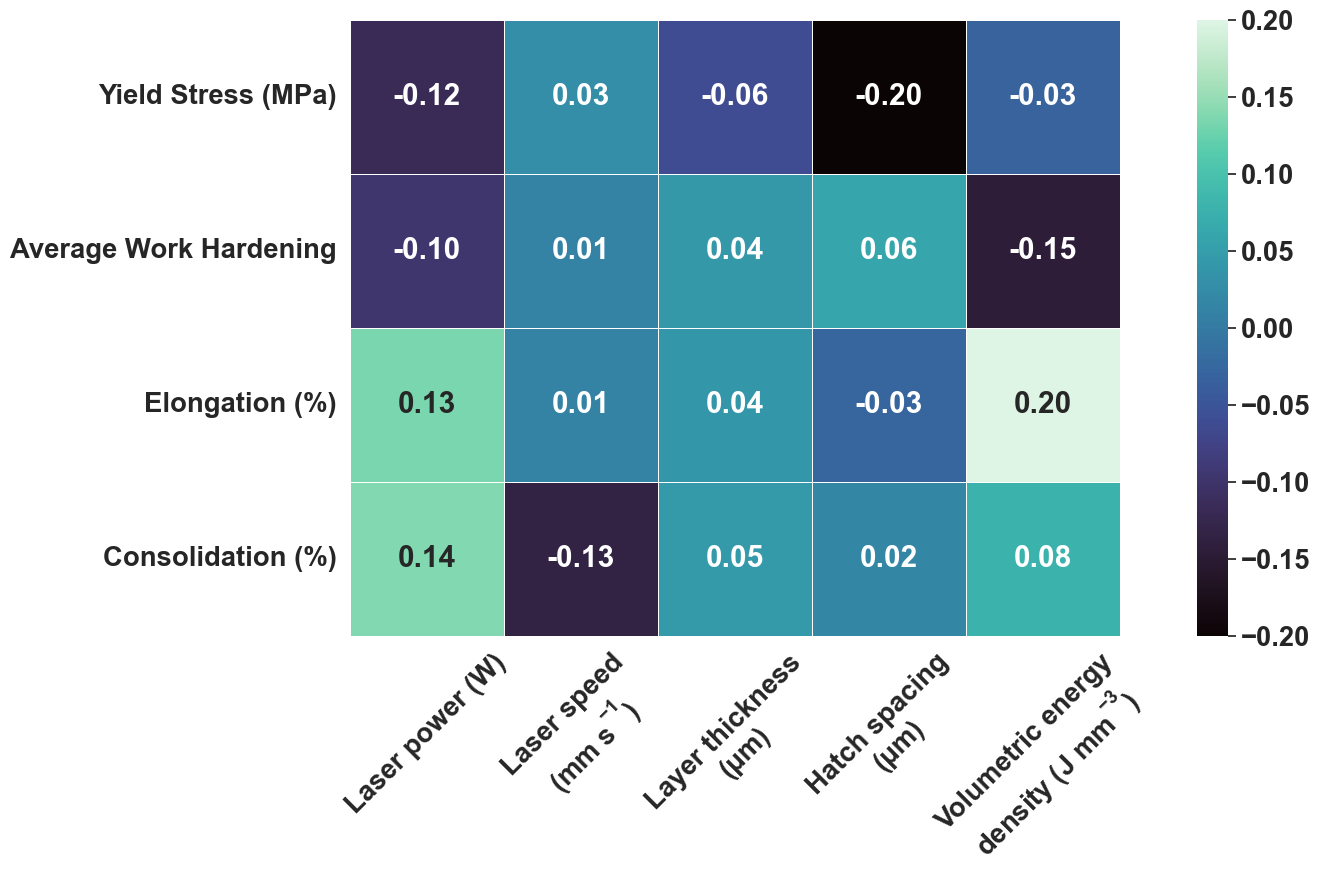}
    \label{fig:Pearsons Output Heatmap}
\end{subfigure}
\caption{\textbf{(a)} Heatmap of Spearman's rank correlation coefficient between processing parameters and output properties, denoting the strength of monotonic relationship between subsequent variable pairs. \textbf{(b)} Heatmap of Pearsons's rank correlation coefficient between processing parameters and output properties, denoting the strength of linearity between subsequent variable pairs.}
\label{fig:Heatmaps}
\end{figure}

Spearman's rank and Pearson's correlation coefficient between the process parameters and relevant outputs are shown in \textbf{Figure \ref{fig:Spearmans Output Heatmap}} and \textbf{Figure \ref{fig:Pearsons Output Heatmap}}, respectively. Spearman's rank has been used as it is more appropriate for heavy-tailed distributions and can be used to uncover monotonic relationships \cite{DouglasC.Montgomery2012IntroductionAnalysis,Hogg2018IntroductionStatistics}. Thus, Spearman's rank is suitable for the obtained dataset, where the majority of distributions for both input parameters (\textbf{Figure \ref{fig:Input Histogram}}) and output properties (\textbf{Figure \ref{fig:Output Histogram}}) are both heavy-tailed. Whereas Pearson's is suitable to examine the linearity between variables. It is highly unlikely that the underlying physics of AM is linear, hence the Pearson's correlation coefficient reflects the non-linearity between two given variables \cite{DouglasC.Montgomery2012IntroductionAnalysis,Hogg2018IntroductionStatistics}. 


Volumetric energy density (VED) is a commonly used metric that consolidates the key process parameters such as power, beam speed, layer thickness, hatch spacing into a unified parameter and use it to optimize the print quality. Therefore, analysis of the correlation between VED with output variables such as consolidation and mechanical properties (reflecting the build quality) was also included in this study to discuss the effectiveness of the use of the VED.
The colorbar besides the heatmap represents the correlation between the corresponding variables, where 1 denotes a perfect positive monotonic/linear relationship and -1 denotes a perfect negative monotonic/linear relationship, thus a value of 0 signifies no correlation. The closer the correlation is to $\pm 1$, the more precise the association between two variables can be explained by their corresponding monotonic/linear relationship \cite{Hauke2011ComparisonData,2016SupplementalData}. However, the range limits on the bar has been scaled to [-0.4, 0.4] for a clearer presentation. 
Both Spearman's (\textbf{Figure \ref{fig:Spearmans Output Heatmap}}) and Pearson's correlation coefficients (\textbf{Figure \ref{fig:Pearsons Output Heatmap}}) indicate that overall laser power has the strongest correlation to the four quality variables, with VED showing similar coefficients but slightly weaker scores.
Where the highest Spearman's correlation coefficient is 0.31 for laser power and elongation, while the highest Pearson's correlation coefficient is 0.20 for the correlation between layer thickness and VED with elongation. 
Nevertheless, with no coefficients exceeding $\pm$0.5, the low magnitude of the calculated coefficients suggests weak correlations between the processing parameters and the output properties. This is clearly not correct and highlights detrimental implication of biases within the data and/or the non-monotonicity in the correlationship.
One major bias is as highlighted earlier regarding the availability of full spectrum of data in literature (\textbf{Figure \ref{fig:Output Histogram}} - \textbf{Figure \ref{fig:Input Histogram}}). Majority of literature data only reported the consolidation and mechanical properties corresponding to optimized print parameters. Authors only published (or get their studies published) data of high quality builds while the data corresponding to low quality were not published. This bias is particularly shown in the consolidation reports: over 80\% of reported consolidation values are over 95\%, \textbf{Figure \ref{fig:Output Histogram}}. All available data of print parameters only correspond to a narrow range of values (limited to high consolidation and optimal mechanical properties). The lack of data that show strong effects of process parameters on build quality outside of the narrow range of optimized values causes the available data to fail at reflecting a strong correlation between the print parameter variables and output variables, explaining  the low values for the Spearman's and Pearson's coefficients. Such a bias in only reporting optimized parameters makes the use of reported data fail to accurately capture the full spectrum of the process - property relationships \cite{Gordon2020DefectManufacturing}.
Another reason is that the complex physics governing the melting and solidification of LPBF process can't be captured with a single one-to-one monotonic correlation between individual variables. This suggests a metric (e.g. VED) involving all parameters simultaneously should be capable of capturing the correlation of input - output variables. Surprisingly, the correlations between VED and the quality variables were quite similar to those of the laser power. 
It is important to note that the correlation values calculated in \textbf{Figure \ref{fig:Heatmaps}} may be affected by the presence of multicollinearity among the processing parameters. This is true for AM processes in which multiple parameters are tuned to achieve optimized quality. The multicollinearity is also reflected in a fact that the Pearson's correlation values were quite low \cite{Hauke2011ComparisonData,2016SupplementalData}.

The variance inflation factor (VIF) is presented in \textbf{Table \ref{tab:VIF}} to examine the degree of multicollinearity. VIF suggests that all considered parameters are highly multicollinear with the layer thickness and hatch spacing has the highest VIF, suggesting that the value choice of one (or both) of these two process parameters is dependent on the choice of other parameters. 
It is likely that the collinear dependence is extrinsic and engineered by printer users in optimizing the build quality. Laser power, speed and other parameters are often tied to one another to optimize the print quality \cite{Gordon2020DefectManufacturing}.
To examine the correlation of an individual input parameter with output variables, it is necessary to have clean data in which only an input parameter is varied while the other input parameters are fixed. Unfortunately, such clean data is not publicly available in published literature.
Most common method to address multicollinearity involves removing features with high VIF. However, as the objective is to understand how process parameters affect the build quality, removal of any features is avoided.
Thus, algorithms immune to multicollinearity will be used to train ML models, such as neural networks and tree-based algorithms. Where neural networks are not affected due to the overparameterization of coefficients or weights at each layer of the network, rendering the inflated regression coefficients redundant \cite{DeVeaux1994}. Whereas tree-based algorithms selects single features at a time when splitting the tree in a forward-stagewise manner, improving the model as demonstrated by Hastie et al. \cite{hastie01statisticallearning}.


\def\arraystretch{1.35}
{\setlength{\tabcolsep}{2em}
\begin{table}[h!]
  \begin{center}
    \begin{tabular}{c c}
      \hline
      \textbf{Variable} & \textbf{VIF}\\ 
      \hline
      Laser power (W) & 4.78 \\
      Laser speed (mm/s) & 2.92 \\
      Layer thickness ($\mu$m) & 8.32 \\
      Hatch spacing ($\mu$m) & 8.56 \\
      \hline
    \end{tabular}
    \caption{Multicollinearity detection of processing parameters with variance inflation factor (VIF).}
    \label{tab:VIF}
  \end{center}
\end{table}


\subsection{Principal Component Analysis (PCA)}
\label{sec:ch3_subsec22}

VED is often used in literature to reducing the dimensionality in the relationship between the quality of a build and process parameters. To further evaluate the efficacy of reducing the dimensionality in the data (in particular, the use of VED as a parameter to optimize print quality), PCA has been undertaken to reduce the dimensionality of all the process parameters into two variables for visualization. Where VED is, in effect, also a form a dimensionality reduction, thus a comparison would show the efficacy between both VED and the generated principal components ability to capture the correlation between process parameters and material properties. The PCA transformation used for this study was performed as follows


\begin{equation}
\label{eq:PCA Matrix}
\centering
\begin{bmatrix}
 0.5238 & -0.0316 & 0.5966 & 0.6072 \\
 0.4325 & 0.8685 & -0.1185 & -0.2114 \\ 
\end{bmatrix}
\times
\begin{bmatrix}
 P_{1} & P_{2} & \cdots & P_{n} \\
 V_{1} & V_{2} & \cdots & V_{n} \\ 
 T_{1} & T_{2} & \cdots & T_{n} \\ 
 S_{1} & S_{2} & \cdots & S_{n} 
\end{bmatrix}
=
\begin{bmatrix}
 PC_{1,1} & PC_{1,2} & \cdots & PC_{1,n} \\
 PC_{2,1} & PC_{2,2} & \cdots & PC_{2,n} \\ 
\end{bmatrix}
\end{equation}



such that the product of the eigenvectors and P, V, T and S (power, scanning speed, layer thickness and hatch spacing respectively) is used to calculate the principal components $PC_{1}$ and $PC_{2}$. The eigenvector is generated by performing eigen decomposition on the covariance matrix of the dataset. 
Only two eigenvectors with the highest eigenvalues, $PC_{1}$ and $PC_{2}$ are retained, as these correspond to the two eigenvectors which capture the highest amount of variance in the data.
The transformed space of both PC is represented in the biplot shown in \textbf{Figure \ref{fig:PCA Biplot}}. 
The length of the arrow depicts the strength of an individual process parameter with respect to its PC direction, whereas the angle represents the contribution of the process parameter to a PC: e.g., if a parameter is parallel to $PC_{1}$, it contributes only to this component \cite{James2021AnEdition,Abdi2010PrincipalAnalysis,Jollife2016PrincipalDevelopments}. 
Thus, the transformed space suggests that layer thickness and hatch spacing contributes mostly to $PC_{1}$, whereas laser speed mostly contributes to $PC_{2}$, with laser power contributing to both equally.
The angle of the arrows shows that the most influential variables for the construction of $PC_{1}$ are layer thickness, hatch spacing and the laser power, whereas laser speed primarily contributes to $PC_{2}$. Although the arrow lengths suggests laser speed has the greatest influence in generating a principal component, the biplot shows laser power correlates to all other process parameters studied. It should note that a principal component analysis can capture the correlation between constituent variables. The layer thickness and hatch spacing are very similar in creating the PC1, this suggests the two parameters are highly correlated with each other, where laser power is equally correlated to the laser speed, and the hatch spacing/layer thickness. Such correlation is consistent with the VIF values \textbf{Table \ref{tab:VIF}} in which the VIFs of layer thickness and hatch spacing are almost the same while the VIF of the laser power is in between that of laser speed and those of layer thickness and hatch spacing.
This implies that laser power is used as the main parameter to balance with the adjustment of the other parameters on optimizing the quality build.


\begin{figure}[h]
\centering
\includegraphics[width = 0.45\hsize]{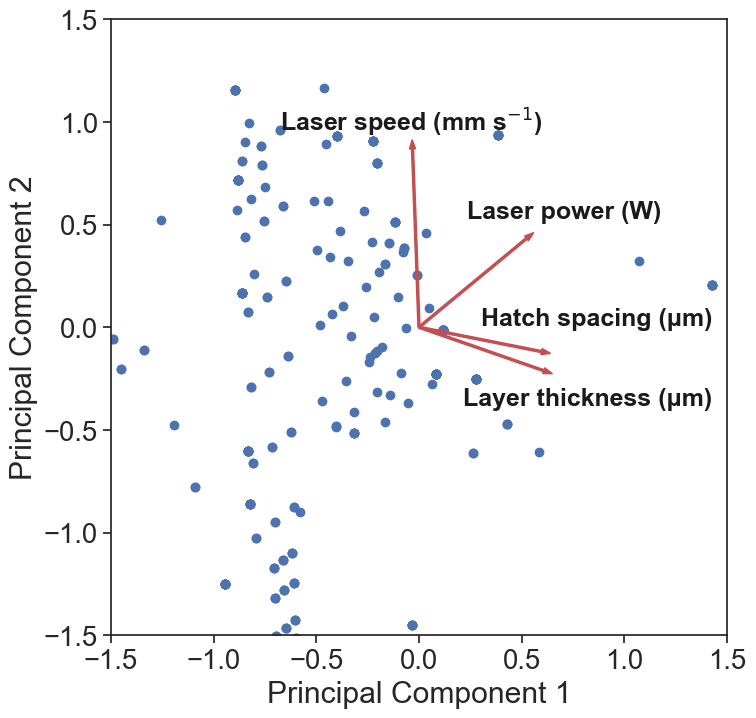}
\caption{PCA biplot representing the contribution and relationship between the processing parameters and both dimensionally-reduced principal components. The scatter plot showcases the projection of original data onto the reduced dimensional space.}
\label{fig:PCA Biplot}
\end{figure}


\textbf{Figure \ref{fig:PCA Spearman}} displays the Spearman's rank heatmap for the two PCs. The generated map shows that the $PC_{1}$ has stronger monotonic correlation with the quality variables (apart from the work hardening) than VED, hence is more effective at capturing the relationship between process parameters and quality. However, $PC_{2}$ performs a lot worse than VED, suggesting that $PC_{1}$ is capable of capturing the majority of the correlation between the process and properties. This is due to the process of PCA transformation: To maximise the captured variance in one direction, the other is reduced in the process \cite{James2021AnEdition,Abdi2010PrincipalAnalysis,Jollife2016PrincipalDevelopments}.
The Spearman's and Pearson's rank heatmaps (\textbf{Figure \ref{fig:PCA Spearman} - \ref{fig:PCA Pearson}}) show the PC's better ability in capturing the correlation in comparison to VED. In particular, $PC_{1}$ outperforms the VED at capturing the correlation of YS, elongation and consolidation.
Nevertheless, despite popular belief, the results of the correlation for $PC_{1}$ questions the suitability of VED as a metric to optimize the print quality. However, all the correlation values of PC1 are still relatively low, all being less than 0.5. This may imply potential issues in ML training as the machine may struggle with uncovering underlying patterns and correlation between processing parameters and properties \cite{James2021AnEdition}.


\begin{figure}[h!]
\centering
\begin{subfigure}{0.45\linewidth}
    \caption{}
    \includegraphics[width=\linewidth]{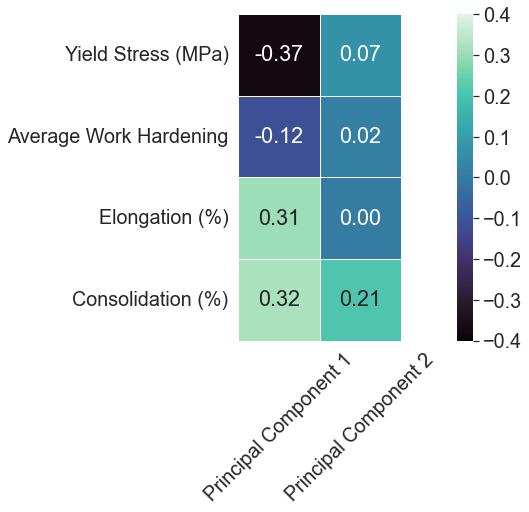}
    \label{fig:PCA Spearman}
\end{subfigure}
\hspace{.05\linewidth}
\begin{subfigure}{0.45\linewidth}
        \caption{}
    \includegraphics[width=\linewidth]{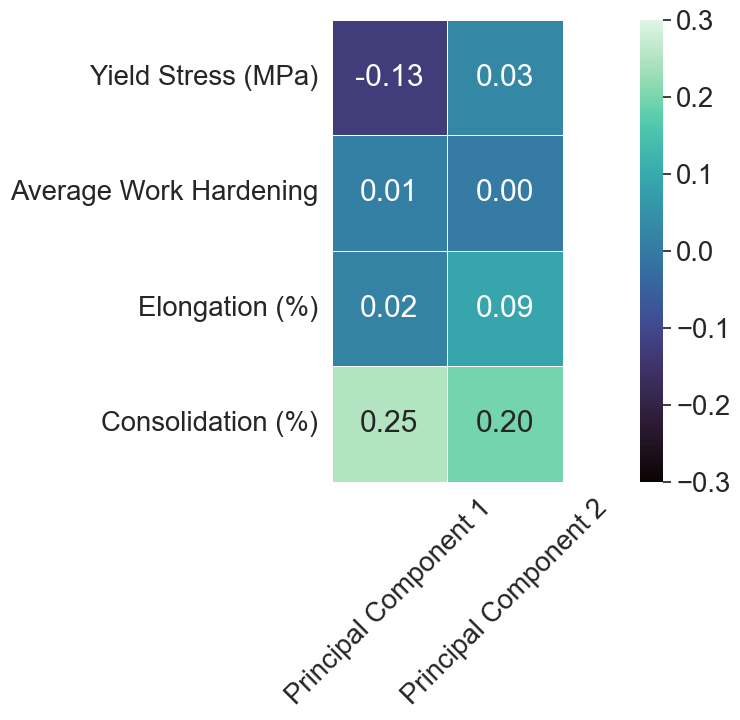}
    \label{fig:PCA Pearson}
\end{subfigure}
\caption{\textbf{(a)} Heatmap of Spearman's rank correlation coefficient between principal component's of processing parameters after dimensionality reduction and output properties, denoting the strength of monotonic relationship between subsequent variable pairs. \textbf{(b)} Heatmap of Pearson's rank correlation coefficient, denoting the strength of linear relationship between subsequent variable pairs.}
\label{fig:PCA}
\end{figure}


\newpage
\subsection{Process Parameter Optimization} 
\label{sec:ch3_subsec24}

The common practice of identifying the process window is based on the consolidation. While achieving high consolidation is important, consolidation is not a single indicator of the material performance in structural application in which more than 90\% of failures is due to mechanical performance, in particular fatigue \cite{suresh_1998}. 
Thus, the aim was to include multiple variables that better reflect the print quality and by analyzing the data across many different groups, the results will yield in a better identification of process window for high quality including consolidation. Therefore, selected processing parameters (laser power and speed) have been optimized with respect to not only the consolidation, but also YS, work-hardening and elongation following the method stated in Section \ref{sec:ch3_subsec13}. However, as the data on fatigue (and creep) is largely missing, these properties have not been considered for the identification of these windows.

The generated optimized maps ($Hmap_2$) for commonly used AM alloys are displayed in \textbf{Figure \ref{fig:Non-treated individual heatmaps}}. 
The process maps ($Hmap_1$) for individual quality variable are provided in Supporting Information \textbf{Figure \ref{fig:316L NT - All Properties} - \ref{fig:Hastelloy NT - All Properties}}. 
The red cross data points depict the collected data, whereas the color of the maps depicts the degree of quality with 1 denoting the highest quality, whereas 0 denoting the lowest quality.
The generated heatmaps considers each alloy individually, as the identification of process parameters is highly dependent on the materials properties. Hence, a single map for all alloys would be obsolete as it wouldn't be an accurate representation of the true processing windows for an individual alloy;
nevertheless, the map considering all materials is provided in Supporting Information \textbf{Figure \ref{fig:NT Summarized Heatmap}} for reference.


\afterpage{
\begin{figure}[H]
{\centering%
\begin{tabular}{@{}lll@{}}
    \begin{subfigure}{0.45\linewidth}
      \caption{}
      \includegraphics[width=0.95\linewidth]{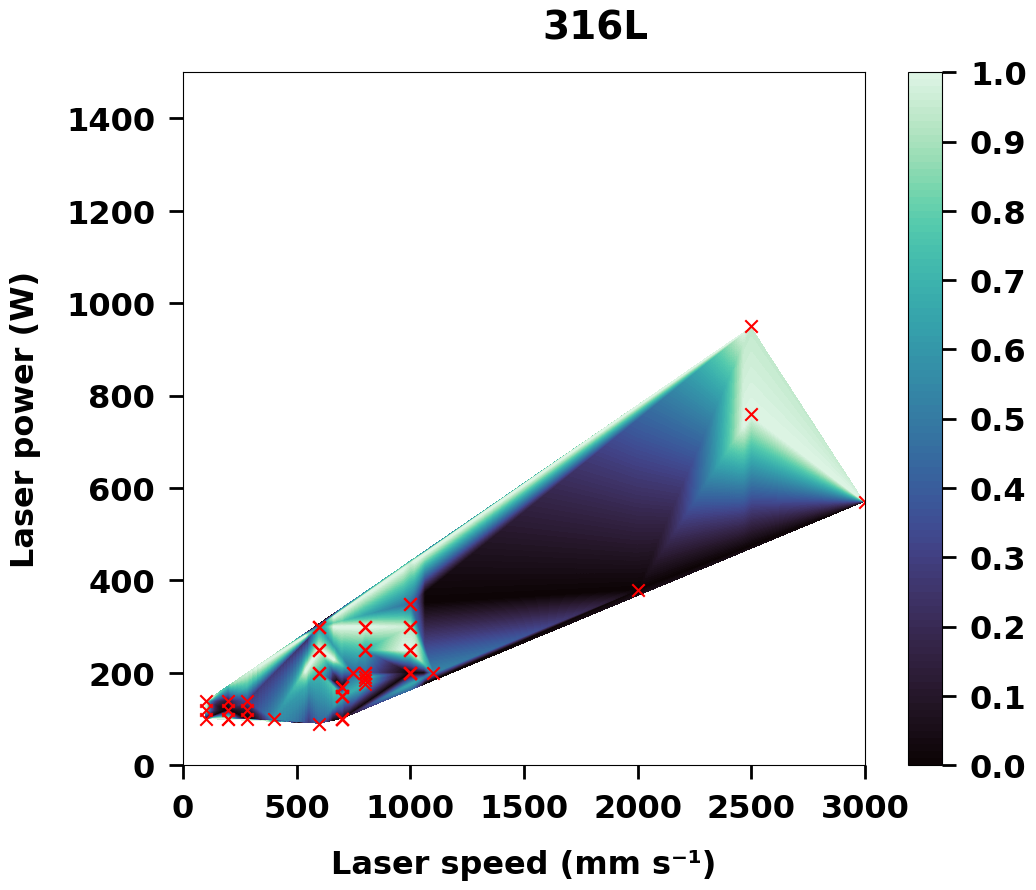} \label{fig:Hmap}
    \end{subfigure} &
    \begin{subfigure}{0.45\linewidth}
      \caption{}
      \includegraphics[width=0.95\linewidth]{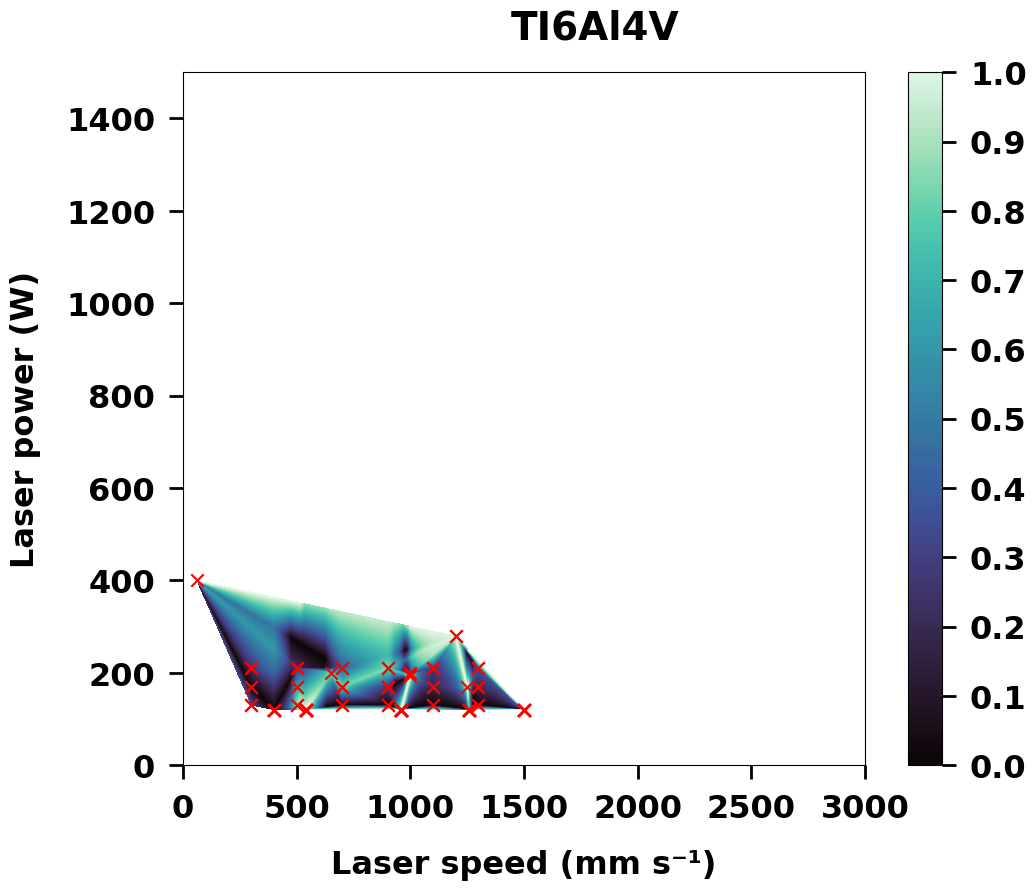} \label{fig:Hmap}
    \end{subfigure} \\
\end{tabular}\par}
{\centering%
\begin{tabular}{@{}lll@{}}
    \begin{subfigure}{0.45\linewidth}
      \caption{}
      \includegraphics[width=0.95\linewidth]{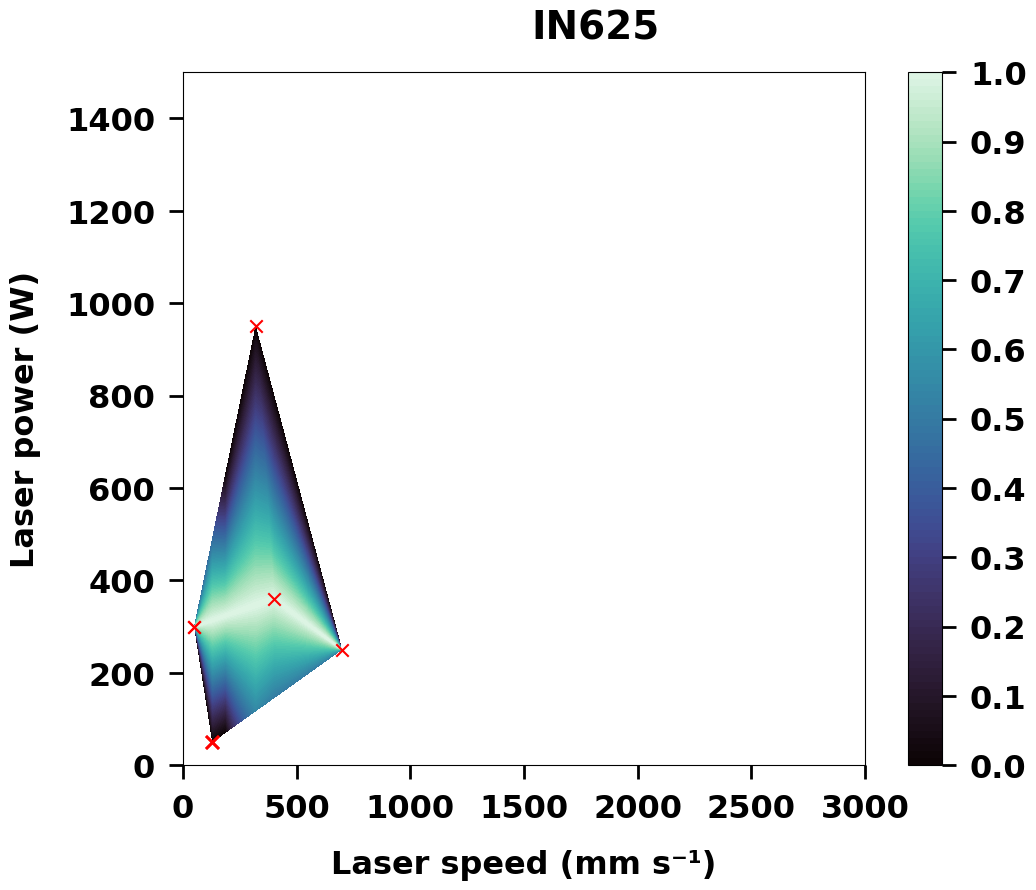} \label{fig:Hmap}
    \end{subfigure} &
    \begin{subfigure}{0.45\linewidth}
      \caption{}
      \includegraphics[width=0.95\linewidth]{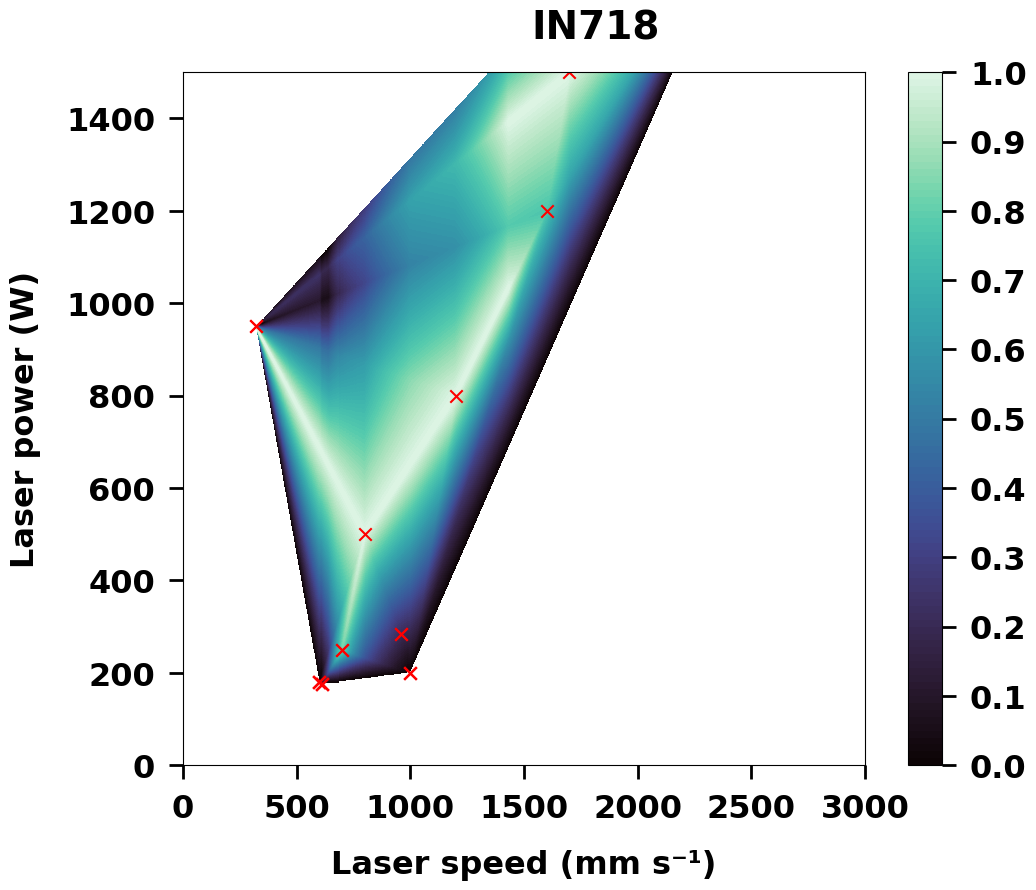} \label{fig:Hmap}
    \end{subfigure} \\
\end{tabular}\par}
{\centering%
\begin{tabular}{@{}lll@{}}
    \begin{subfigure}{0.45\linewidth}
      \caption{}
      \includegraphics[width=0.95\linewidth]{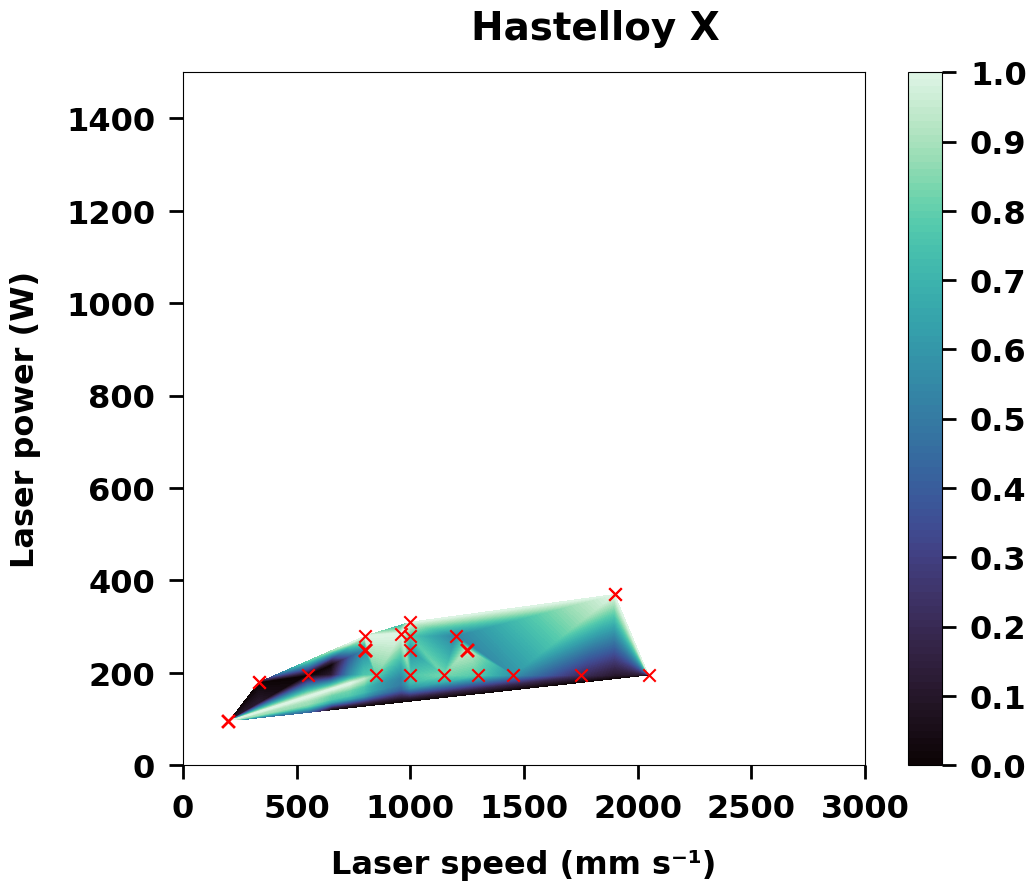} \label{fig:Hmap}
    \end{subfigure}
\end{tabular}\par}
\caption{$Hmap_2$, optimized processing window maps for non-treated \textbf{(a)} 316L, \textbf{(b)} Ti6Al4V, \textbf{(c)} In625, \textbf{(d)} In718 and \textbf{(e)} Hastelloy X. The maps considers the yield strength, average work-hardening, elongation and consolidation when optimizing, where the brighter regions outlines the regions of laser power and laser speed which users are advised to use to obtain better overall print quality.}
\label{fig:Non-treated individual heatmaps}
\end{figure}
\clearpage}


The overall printability of an alloy can be evaluated by the area of the process map for high quality. Such that an alloy with a larger process map can be printed with the high quality and with wider ranges of process parameters, i.e. more printable than another alloy with a smaller process map area. \textbf{Figure \ref{fig:Non-treated individual heatmaps}} suggests 316L and IN718 are most printable amongst all considered alloys. 
Furthermore, the maps suggests that IN625 and Hastelloy X have slightly better printability than Ti6Al4V.
The low printability of Ti6Al4V is likely due to the loss of element (up to 0.9wt\% loss of Al) and low ductility because of martensite and high dislocation densities \cite{Lertthanasarn2021,Mukherjee2016PrintabilityManufacturing}.


\subsection{Machine Learning Performance}
\label{sec:ch3_subsec23}

Following the preprocessing and training procedures stated in Section \ref{sec:ch3_subsec12}, 
the top 20 highest performing hyperparameters were used to train the 7 ML algorithms. The performance of all 140 trained models have been summarized with RMSE as the metric to evaluate the accuracy, \textbf{Figure \ref{fig:Performance Boxplot}}.
The RMSE on the boxplot shows CatBoost and random forest exhibited the highest variability in performance, whereas XGBoost, LightGBM and neural networks demonstrated the least variability. Such that these three algorithms displayed consistent performance among the 20 trained models for each algorithm. However, on average, XGBoost emerged as the best-performing algorithm, surpassing CatBoost by a small margin.
Overall, the performance of all trained models aligns well with existing literature, such that boosting algorithms typically have the best performance, followed by random forest, neural networks, support vector machine and decision tree \cite{Caruana2006}. Notably, the performance of neural network did not agree with this result, likely due to the limited size of the training data, as neural networks typically require large volumes of data to achieve optimal performance \cite{Guo2020}.


\begin{figure}[h]
\centering
\includegraphics[width = 0.75\hsize]{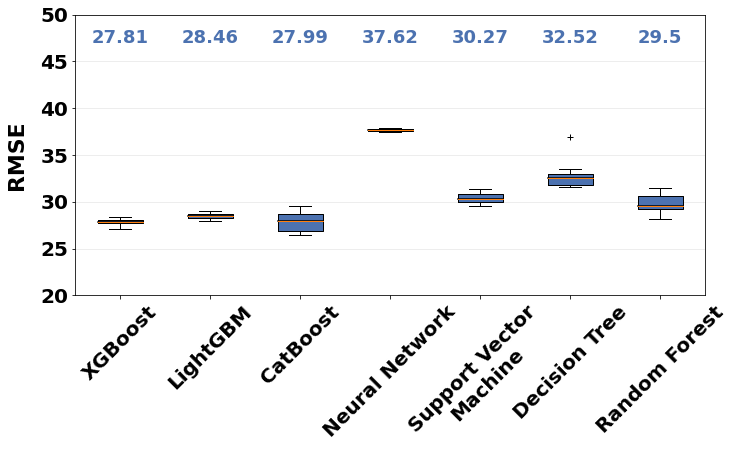}
\caption{Boxplot comparing RMSE performance metric of all trained machine learning models, where the median of the samples is displayed at the top.}
\label{fig:Performance Boxplot}
\end{figure}


To assess the ML performance, the testing dataset has been fed to the best performing XGBoost model to give predictions of YS, average work-hardening, elongation and consolidation. \textbf{Figure \ref{fig:Training}} shows the comparison between the true values obtained from literature and the output predictions from the trained XGBoost model.
The predictions for YS and elongation shows great agreement with the true values, the observation is reinforced by the calculated coefficient of determination ($R^2$) which were 0.91 and 0.83, respectively. The $R^2$ values suggests that 91\% and 83\% of the variance can be explained by the model, for YS and elongation respectively. Notably, the $R^2$ value for YS was the highest in all trained ML models. This high performance of ML on the YS is likely because of the high correlation between YS and the primary dendritic (or cellular) spacing. It is found that the YS is inversely related to the cellular dendritic spacing that is inversely proportional to the cooling rate, which is, in turn, controlled by the process parameters \cite{Piglione2021, Pham2020}. This implies that there is a well defined correlation between the process parameters and YS reflected by the ML good performance, suggesting that ML is capable of reflecting the underlying science of such relationship and such a capability should be improved if there were more studies reporting the cellular spacing.
The XGBoost model provides a reasonable prediction of the work hardening. As the work hardening was calculated based on YS, UTS and elongation, thus the accumulation of errors in predictions is expected to the reduce accuracy of work hardening predictions. However, the high accuracy predictions for YS and elongation suggests the ML's low performance on the hardening was mainly due to the accuracy of predicting UTS.
The ML model performed worst regarding the consolidation with 67\% of the variance not predicted by the model. The low performance of consolidation (\textbf{Figure \ref{fig:Training}\hyperref[fig:Training]{c2}}) highlights the detrimental consequence of the bias in reporting the consolidation values as discussed earlier: The majority of consolidation reports was \(>\)98\% (\textbf{Figure \ref{fig:Output Histogram}}). Such a bias negatively affected the learning of the ML models. Furthermore, because the majority of consolidation data lies within the 90\% - 100\% range, the ML's prediction on the consolidation was highly weighted by the training of ML on the known data of this range. The performance of ML worsens for predicting consolidation further away from the known range (\textbf{Figure \ref{fig:Training}\hyperref[fig:Training]{c1}}). Moreover, the low performance of ML may be related to the difficulty in capturing the stochastic nature of porosity formation in the melting and cooling of AM process. Last but not least, as the property is influenced by the build location/direction due to the build up of residual heat and scanning strategy which were not included in the training of ML models; lowering the prediction accuracy of ML. To address this limitation, these relevant process parameters should ideally be incorporated into the training of ML model. However, this was not implemented due to insufficient data availability concerning these factors.




Although the model considers whether the sample has been heat treated or HIP by means of \textit{One Hot Encoding}, the results suggests no clear distinction in affecting the accuracy of predictions between treated and non treated samples; however, over time with more collected data, it is likely that the these observations may differ.
Nevertheless, the results shown in \textbf{Figure \ref{fig:Training}} suggests that the trained ML model has some credit in characterizing the relationship between process parameters and the output properties considered, in particular for YS and elongation.
Furthermore, the difference in performance between simple and complex ML models such as non-boosted and boosted algorithms only yielded in a minor increase in performance, suggesting that the biggest underlying factor is the quality of data. Therefore, getting high quality data with minimal biases would significantly improve the performance of ML models.

\newpage

\begin{figure}[H]
{\centering%
\begin{tabular}{@{}lll@{}}
    \begin{subfigure}{0.4\linewidth}
      \caption{}
      \includegraphics[width=0.95\linewidth]{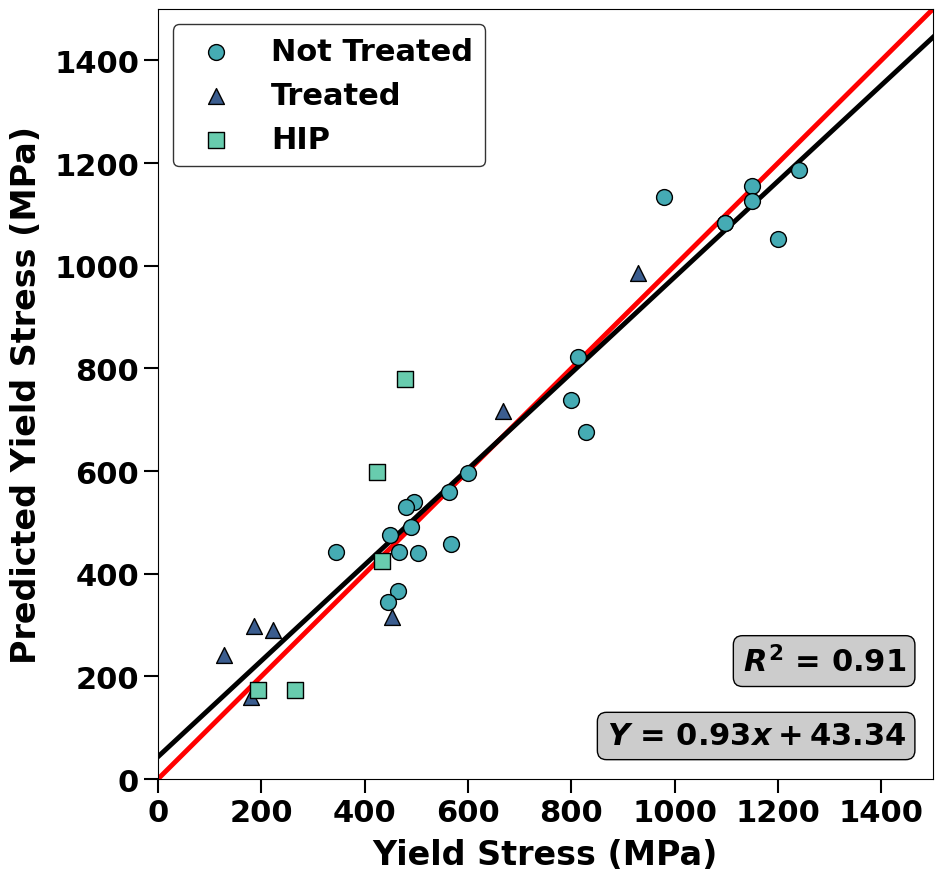} \label{fig:pred YS}
    \end{subfigure} &
    \begin{subfigure}{0.4\linewidth}
      \caption{}
      \includegraphics[width=0.95\linewidth]{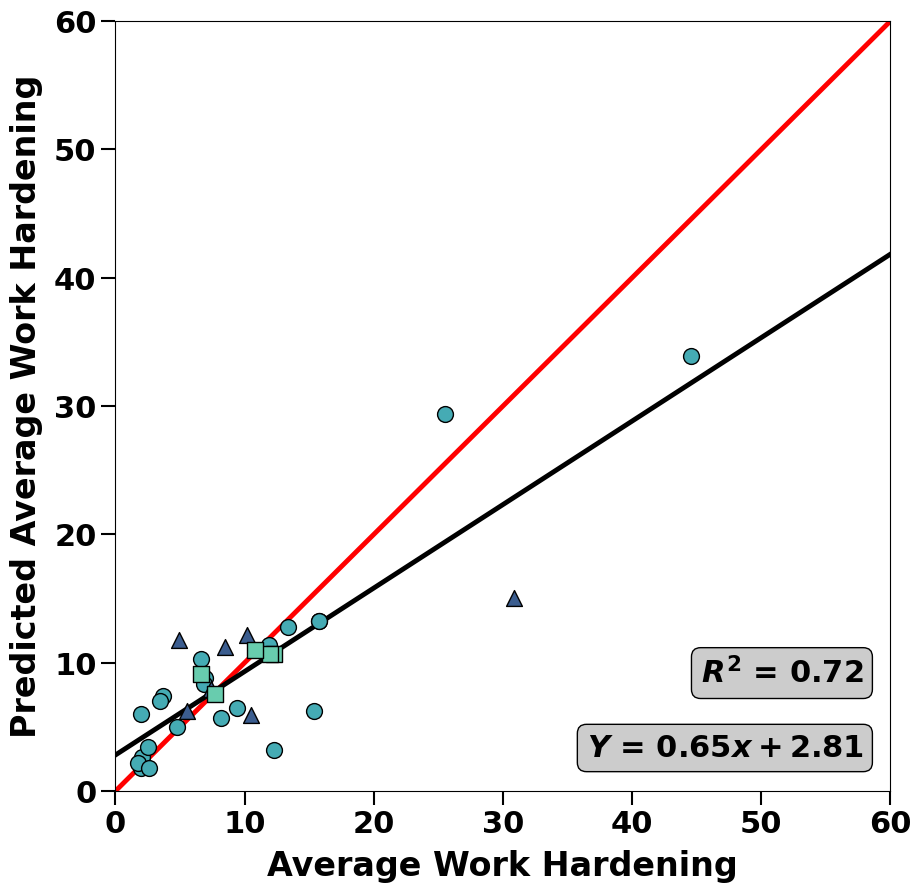} \label{fig:pred Hardening}
    \end{subfigure} \\
\end{tabular}\par}
{\centering%
\begin{tabular}{@{}lll@{}}
    \begin{subfigure}{0.4\linewidth}
      \caption*{(c1)}
      \label{fig:pred Consolidation full}
      \includegraphics[width=0.95\linewidth]{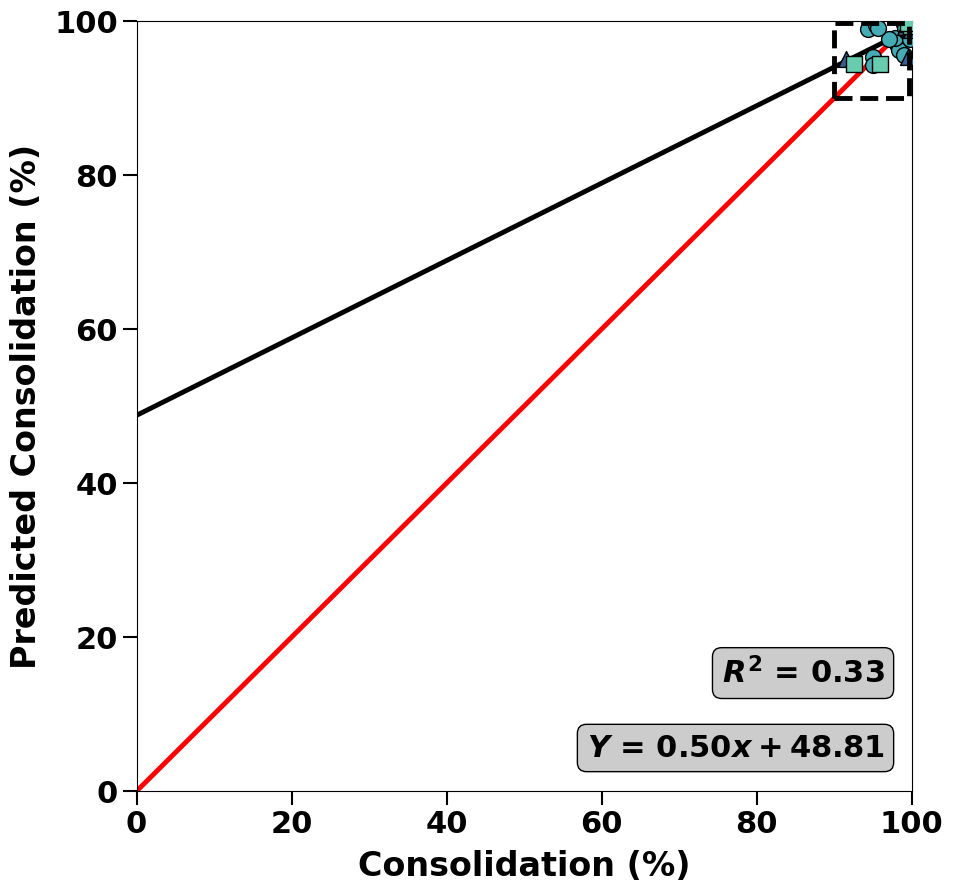} 
    \end{subfigure} &
    \begin{subfigure}{0.4\linewidth}
      \caption*{(c2)}
      \label{fig:pred Consolidation zoom} 
      \includegraphics[width=0.95\linewidth]{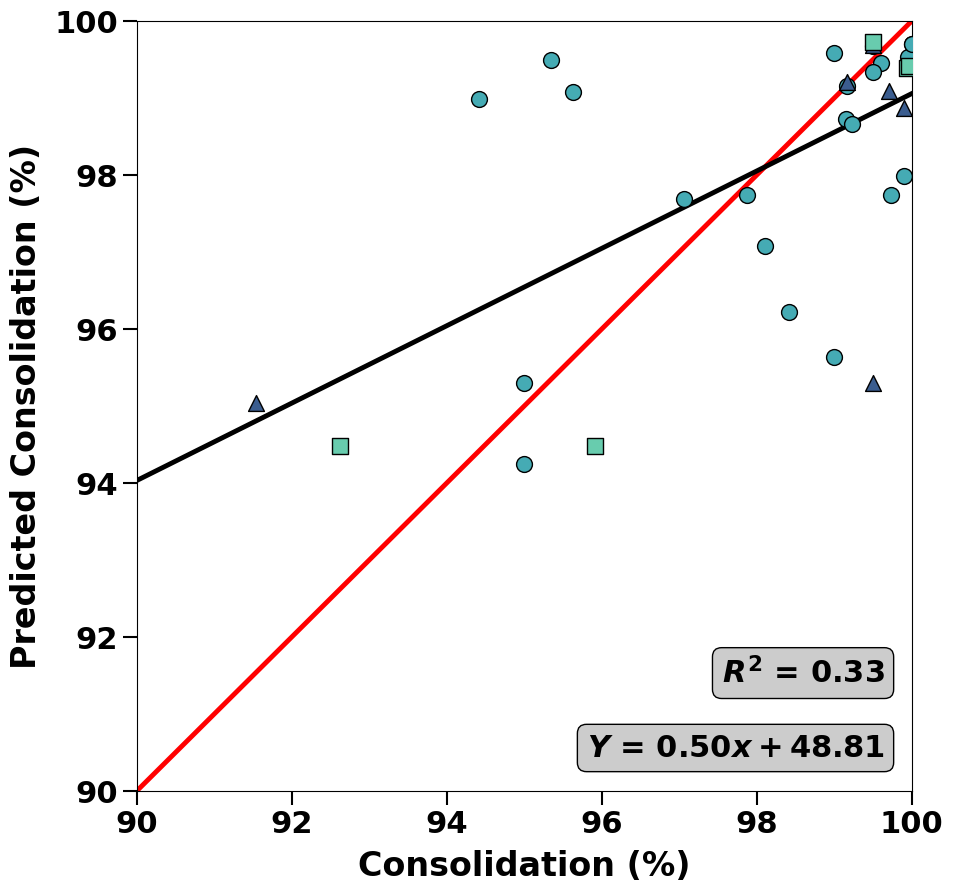} 
    \end{subfigure} \\
\end{tabular}\par}
{\centering%
\begin{tabular}{@{}lll@{}}
    \begin{subfigure}{0.4\linewidth}
      \caption*{(d)}
      \includegraphics[width=0.95\linewidth]{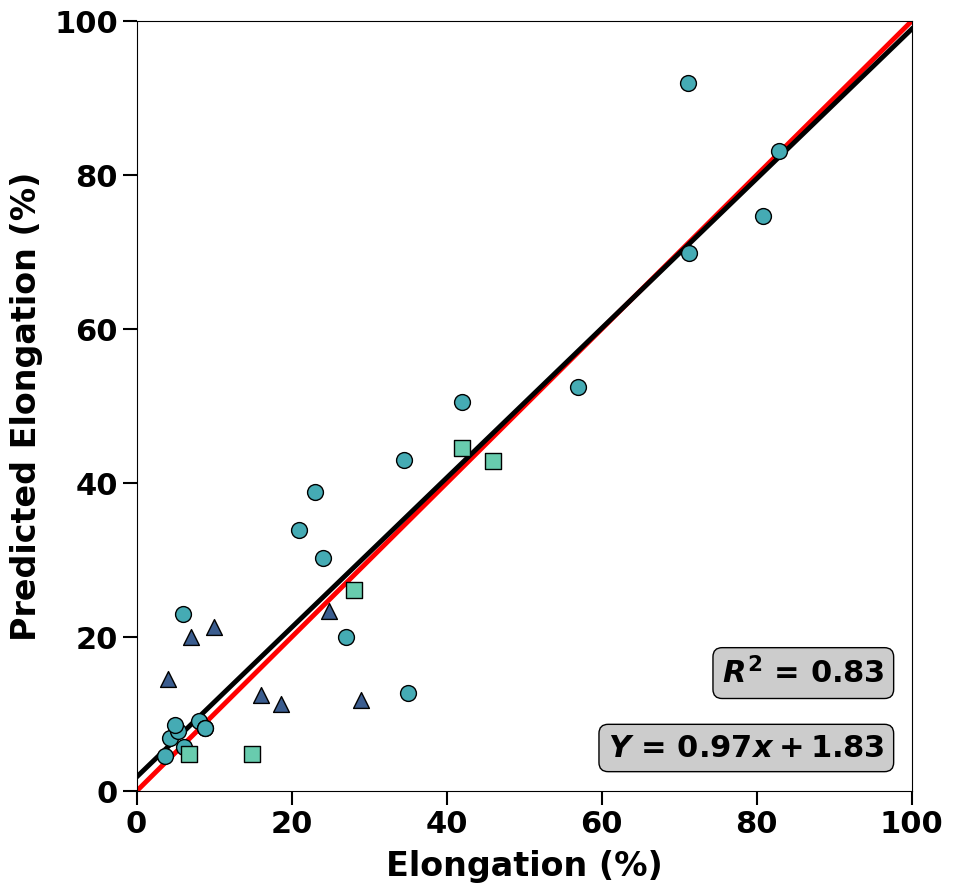} \label{fig:pred Elongation}
    \end{subfigure}
\end{tabular}\par}
\caption{Results of XGBoost model for predicted output properties against experimentally measured values collected in literature for all alloys.
The predicted output properties considered are \textbf{(a)} yield stress, \textbf{(b)} average work hardening, \textbf{(c1)} consolidation, \textbf{(c2)} zoom in view of consolidation and \textbf{(d)} elongation.
The red line depicts a perfect 1 to 1 fit, whereas the black line represents the line of regression obtained from the XGBoost model.}
\label{fig:Training}
\end{figure}}


\newpage
\subsection{Sensitivity Analysis}
\label{sec:ch3_subsec25}

\begin{figure}[h!]
\centering
\includegraphics[width = 0.9\textwidth, keepaspectratio]{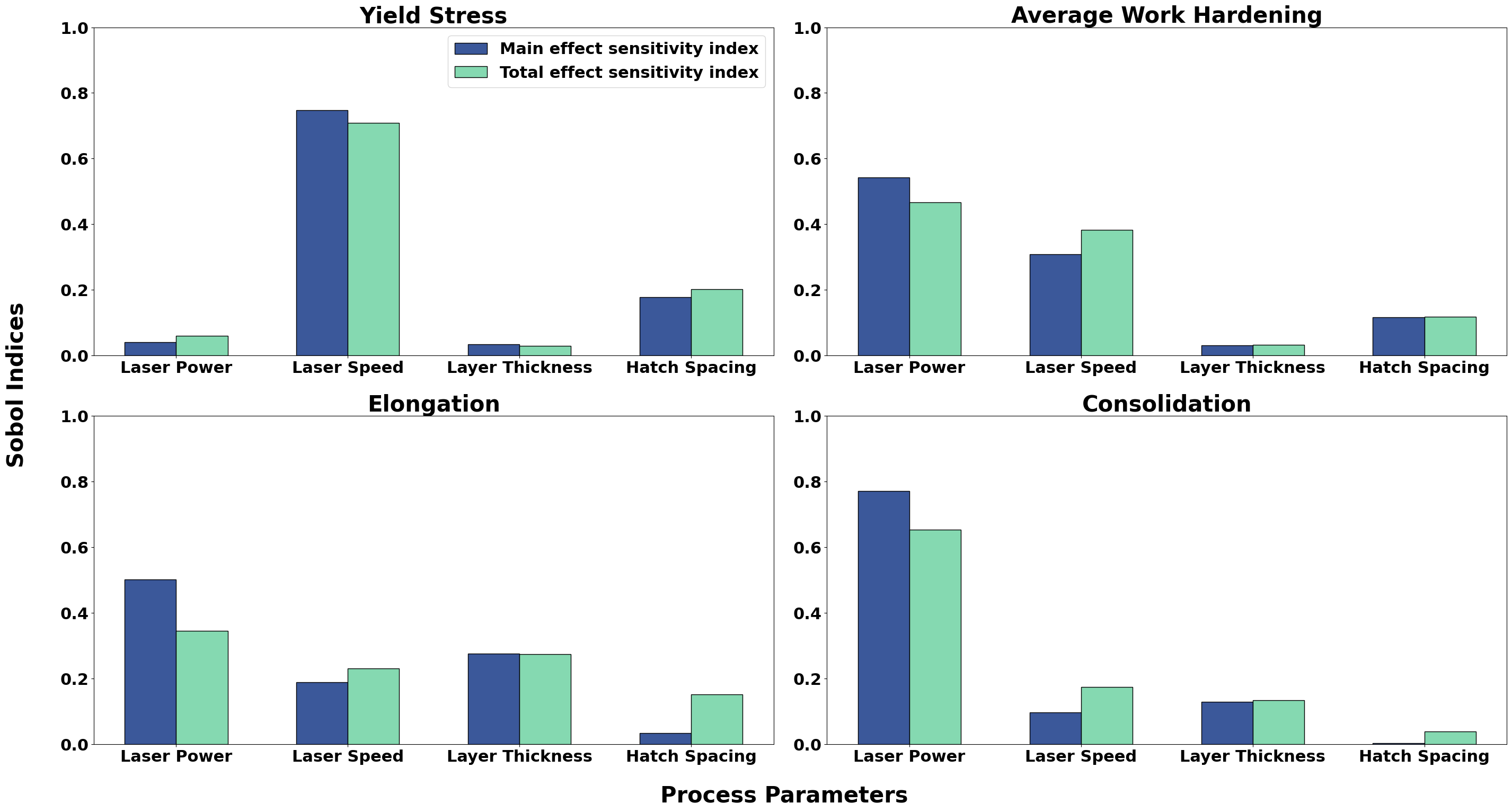}
\caption{Bar plots comparing the main effect sensitivity index and the total effect sensitivity index for the XGBoost model. The main effect sensitivity index measures the influence of an individual process parameter on a given output property, without considering its interactions with other process parameters. Whereas total effect sensitivity index measures the overall influence of an individual process parameter on a given output property, including both the main effect sensitivity and effects through its interactions with other process parameters.}
\label{fig:XGBoost SA}
\end{figure}

Sensitivity analysis has been used to examine the trained ML models ability to account for important influence of AM process parameters in the quality. The best performing XGBoost and CatBoost model have been used for the analysis.
The Sobol indices for the predicted properties by the XGBoost model is displayed in \textbf{Figure \ref{fig:XGBoost SA}}.
The values have been normalised such that the sum of the main effect sensitivity index is equal to 1, as does the sum of the total effect sensitivity index.
The results shows that overall, laser power and speed are the most influential process parameters for all the investigated quality variables. 
However the results show that XGBoost was not able to reflect the underlying science of the process - mechanical property relations. For example, while the XGBoost model shows the laser power was seen highly influential in consolidation, the influence of laser speed was very weak. This is not consistent with a fact that both the laser power and speed are often used to optimize the consolidation \cite{Gordon2020DefectManufacturing}.
In addition, the ML model shows a large influence of laser speed in YS, but not the laser power despite the two key parameter are used in controlling the thermal condition, in particular the cooling rate that governs the spacing of primary dendrites or cells \cite{Pham2020,Bikas2016AdditiveReview,Kotadia2021AProperties,Liu2022AdditiveControl,Bajaj2020SteelsProperties}.


\begin{figure}[h!]
\centering
\includegraphics[width = 0.9\textwidth, keepaspectratio]{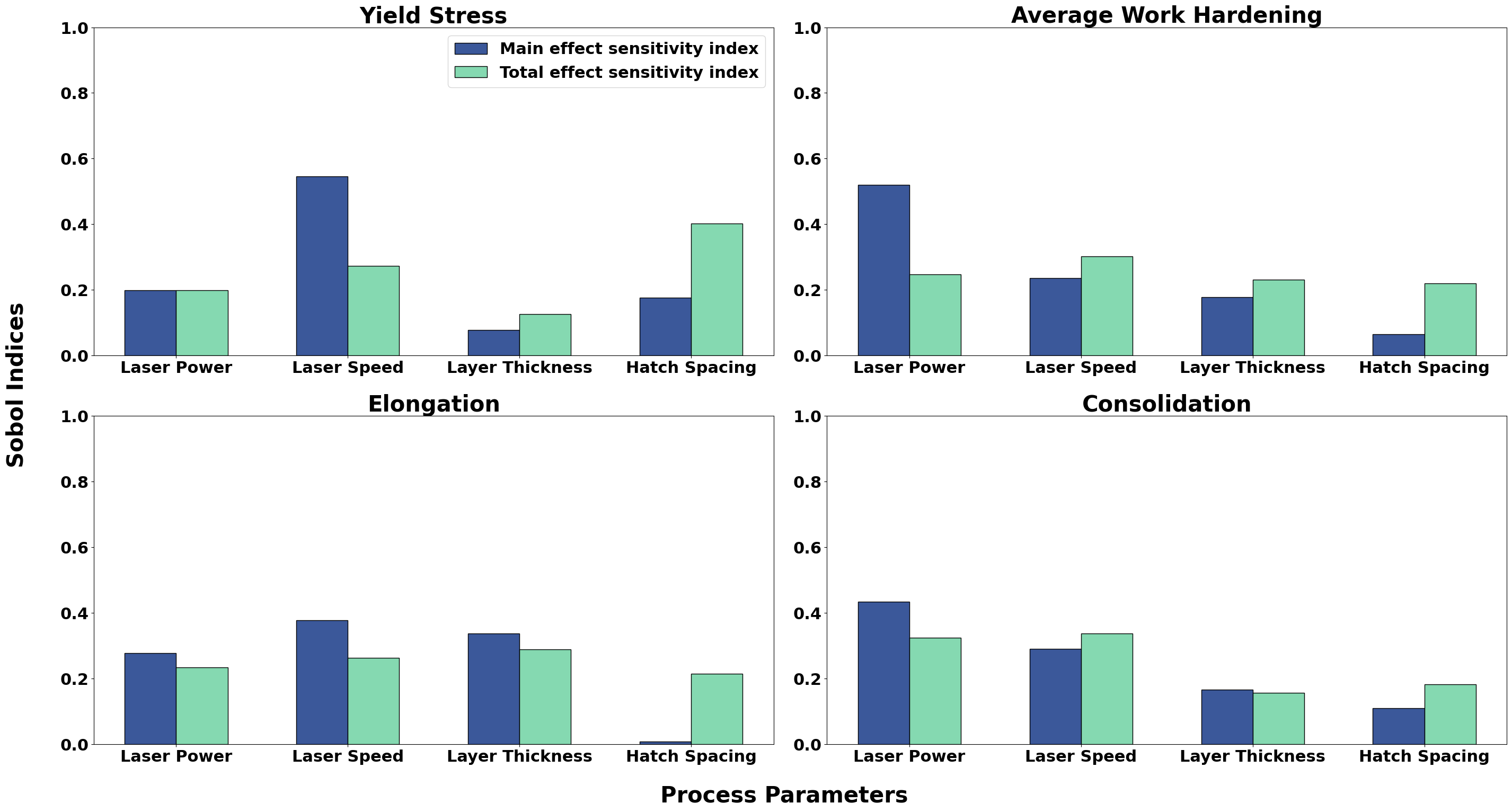}
\caption{Bar plots comparing the main effect sensitivity index and the total effect sensitivity index for the CatBoost model. The main effect sensitivity index measures the influence of an individual process parameter on a given output property, without considering its interactions with other process parameters. Whereas total effect sensitivity index measures the overall influence of an individual process parameter on a given output property, including both the main effect sensitivity and effects through its interactions with other process parameters.}
\label{fig:CatBoost SA}
\end{figure}

Interestingly, the Sobol indices calculated by the CatBoost model (\textbf{Figure \ref{fig:CatBoost SA}}) reflects the known underlying science well.
This can be seen with the Sobol indices for the yield stress, as laser speed and power is shown to be the two most influential process parameters. This aligns well with the correlation between YS with speed and power with the cooling rate, hence YS as discussed earlier. However, hatch spacing is also shown to have high impact, especially with combined interactions as indicated by the total effect index.
Although the ML's prediction for consolidation was suboptimal (\textbf{Figure \ref{fig:Training}}), the calculated Sobol indices reflect the roles of process parameters in the main mechanisms of porosity formation. As process induced pores, such as keyhole pores resulting from excessive power density, and lack of fusion pores caused by insufficient molten metal due to inadequate energy density, i.e. dependent on melt pool geometry, layer thickness and hatch spacing \cite{Sames2016,Pham2020,Gordon2020DefectManufacturing}. Hence, it is anticipated that the influence from processing parameters on consolidation are expected to be approximately equal - this is reflected well in (\textbf{Figure \ref{fig:CatBoost SA}}). 
The mechanism of work hardening relies on interactions of dislocations between themselves and with other crystallographic features. In additive manufacturing conditions, alloys often consist of high dislocation density regions at the cellular (or dendritic) boundaries. Consequently, finer cells (or dendrites) lead to increased interactions between mobile dislocations and immobile dislocations at the dislocation-rich regions. Thus, the primary factors influencing this property are laser power and speed, as they dictate the cooling rate, hence the cell (dendrite) spacing \cite{dieter1988mechanical, Pham2017}.
Whereas elongation is dependent on both consolidation and work hardening. As such, it is expected to exhibit a similar influence of process parameters as the as the combined effect of these two properties.
The analysis also revealed that interactions between process parameters can significantly affect the studied quality variables, indicating the presence of synergistic effects. This can be seen with hatch spacing as the total effect index is always greater than the main effect index. This suggests that although hatch spacing alone has weak influence on properties, the interaction of hatch spacing with other process parameters was found to have substantial impact on properties.


\newpage
\section{Conclusion}

A considerably large literature data for metal additive manufacturing (AM) was created in this study. A comprehensive and in-depth examination of the data highlights major biases and limitation of the literature data, limiting the understanding of the process, microstructure, mechanical property (PMP) relationship: (1) Most studies only reported consolidation, with almost three times more reports than mechanical properties such as yield stress and elongation. (2) Most literature data obtained were in optimized or near optimized conditions, with 84\% of the consolidation data reported above values of 95\%. Correlation analysis shows that this bias limits the literature data in revealing the strong correlation between process parameters and quality variables such as the consolidation and mechanical properties. (3) significantly lack of quantitative data on microstructure such as the spacing of primary dendrites or cells.

Meta-analyses of the collected were done, showing weak correlation between the process parameters (i.e. input) including the volumetric energy density (VED), with the consolidation and mechanical properties (i.e. output). Such weak correlation is likely due to (1) the stated biases and the (2) the correlation between input process parameters and quality variables are non-monotonic with high multicollinearity. The hatch spacing and layer thickness are found to be most collinear, reflecting a common practice in AM process identification: the values of these two parameters are often balanced by tuning the beam power and beam speed.
While the correlation analysis and study of ML performances demonstrate potential for data-driven approaches for metal additive manufacturing, the quality of the dataset hinders the results due to current reporting biases and practices.

The bias (1) reflects another common practice (in AM publications): the identification of process maps is commonly based only on consolidation. Such bias poises a serious limitation on the optimization of parameters because the quality of an AM build is ultimately governed by mechanical properties such as yield strength, elongation and hardening. The significant data enables us to identify the process map on the basis of not only consolidation, but also yield stress, elongation and work hardening. The process map identification results show that amongst all the alloys considered in this study, 316L and Inconel 718 are the most printable alloys followed by Hastelloy X and Inconel 625, where the least printable alloy is Ti6Al4V.

The present study also investigates dimensionality reduction of processing parameters using principal component analysis. The dimensionality reduction was compared to VED, a common metric that consolidates multiple processing parameters for process optimization. Two principal components show much stronger correlation between the PC with the output, suggesting an alternate way (in comparison to the VED) to reduce dimension in optimizing the build consolidation and quality.

The bias (2) seriously limits the use of machine learning in learning the full spectrum of the process parameter - consolidation relationship, hence negatively affecting the ML performance in predicting the consolidation. This effect results in a low accuracy of ML predictions for consolidation, and most evidently in the low value range of consolidation. Furthermore, the minor increase in accuracy by using boosting algorithms compared to non-boosting algorithms further suggests the quality of the obtained dataset is the most significant factor in improving the ML models' performance. We are, therefore, calling the AM community to publicly share data of wide spectrums, in particular process parameters producing low and intermediate ranges of consolidation and mechanical properties beyond consolidation or density. Furthermore, increase data availability through open-access and standardize reporting formats to provide easily accessible data. To aid in this effort, an online template for users to contribute data or use the training dataset and code associated with this work are available in the open-source Github repository at \sloppy{\href{https://github.com/RaymondWKWong/MetaAnalysis_MetalAM}{https://github.com/RaymondWKWong/MetaAnalysis$\textunderscore$MetalAM}}. 

Last but not least, due to insufficient data reported for the microstructure and mechanical properties, ML would not able to learn the PMP relationship. Given the inherent correlation between microstructure and mechanical properties (in particular for the long term performance such as fatigue), next significant efforts should be given to generating microstructure data and fatigue.


\newpage

\medskip
\textbf{Acknowledgements} \par 

The authors would like to thank the EPSRC for supporting the research [grant number EP/K503733/1].
R. Wong and M.S. Pham would like to thank Jalal Al-Lami for providing part of Inconel 718 literature data used for this study. M.S. Pham, R. Wong and C.S. Maldonado thank the Imperial College London's support via a Imperial-Nanyang Technological University seed fund.

The views expressed in the article do not necessarily represent the views of the U.S. Department of Energy or the United States Government. Sandia National Laboratories is a multimission laboratory managed and operated by National Technology and Engineering Solutions of Sandia, LLC., a wholly owned subsidiary of Honeywell International, Inc., for the U.S. Department of Energy's National Nuclear Security Administration under contract DE-NA-0003525.


\newpage
\section*{Supporting Information} 

\renewcommand\thefigure{S\arabic{figure}}    
\setcounter{figure}{0} 





\begin{figure}[h!]
\centering
\includegraphics[width = 0.55\hsize]{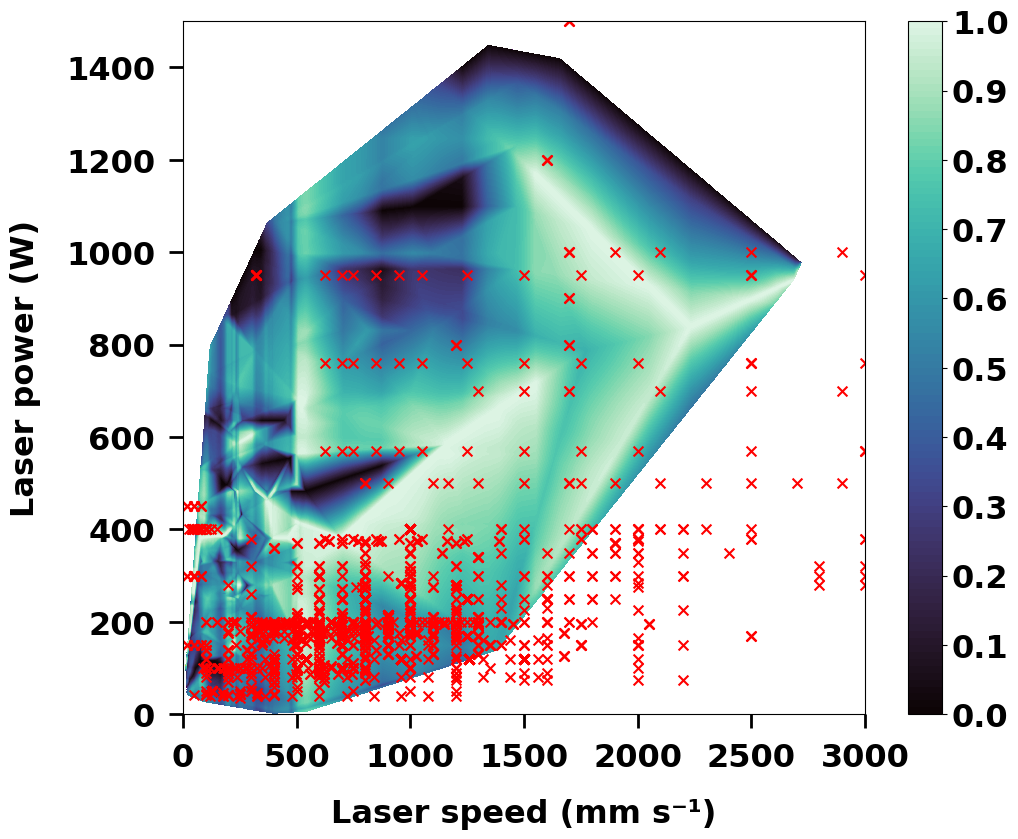}
\caption{$Hmap_2$, optimized processing window maps for the non-treated dataset, where the map considers all materials, optimizing YS, average work-hardening, elongation and consolidation. The brighter regions outlines the regions of laser power and laser speed which users are advised to use to obtain better overall print quality.}
\label{fig:NT Summarized Heatmap}
\end{figure}


\begin{figure}[h!]
\centering
\includegraphics[width = 0.75\hsize]{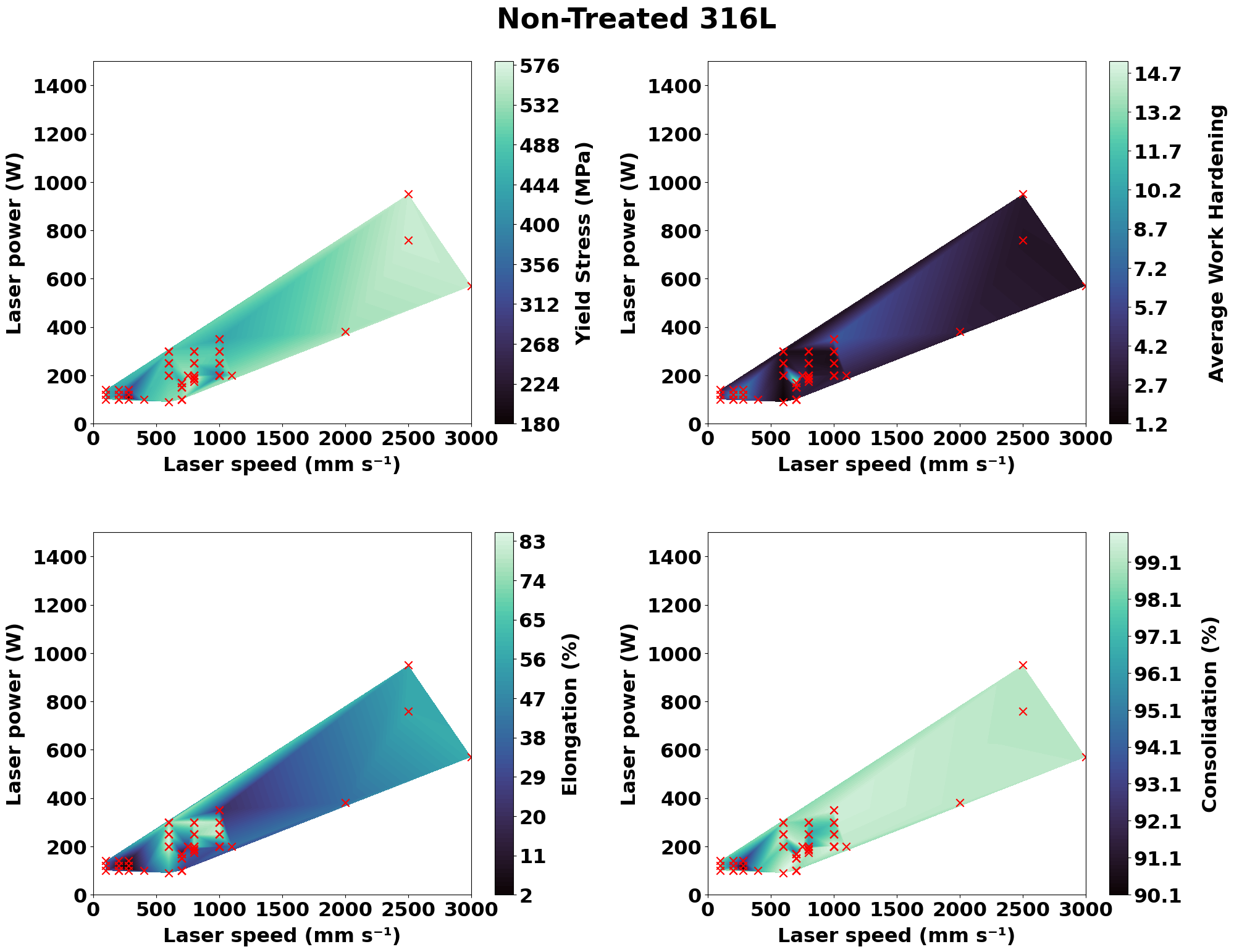}
\caption{Non-treated 316L $Hmap_1$, individual processing window maps for yield strength, average work-hardening, elongation and consolidation.}
\label{fig:316L NT - All Properties}
\end{figure}

\begin{figure}[h!]
\centering
\includegraphics[width = 0.75\hsize]{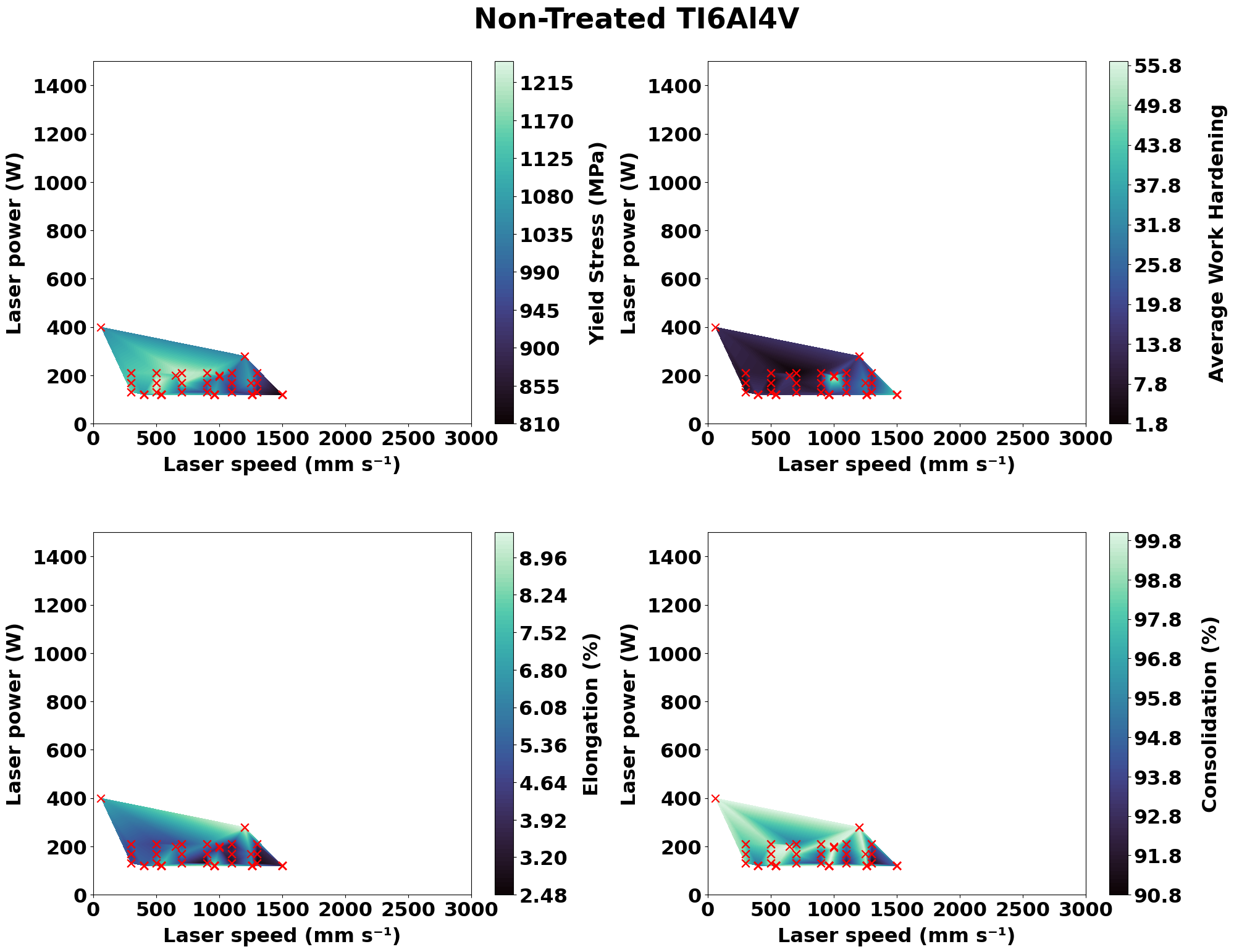}
\caption{Non-treated Ti6Al4V $Hmap_1$, individual processing window maps for yield strength, average work-hardening, elongation and consolidation.}
\label{fig:Ti6Al4V NT - All Properties}
\end{figure}

\begin{figure}[h!]
\centering
\includegraphics[width = 0.75\hsize]{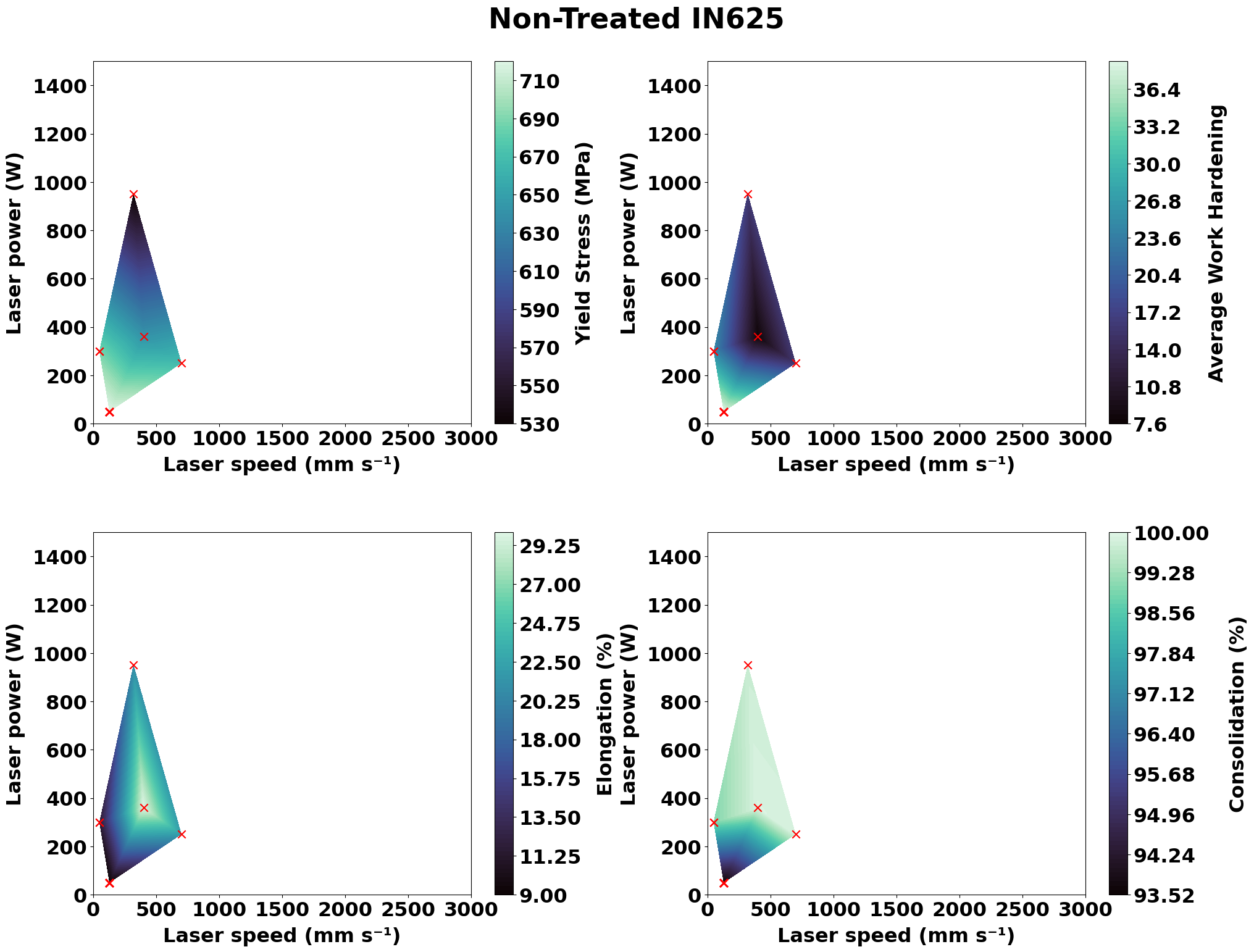}
\caption{Non-treated IN625 $Hmap_1$, individual processing window maps for yield strength, average work-hardening, elongation and consolidation.}
\label{fig:In625 NT - All Properties}
\end{figure}

\begin{figure}[h!]
\centering
\includegraphics[width = 0.75\hsize]{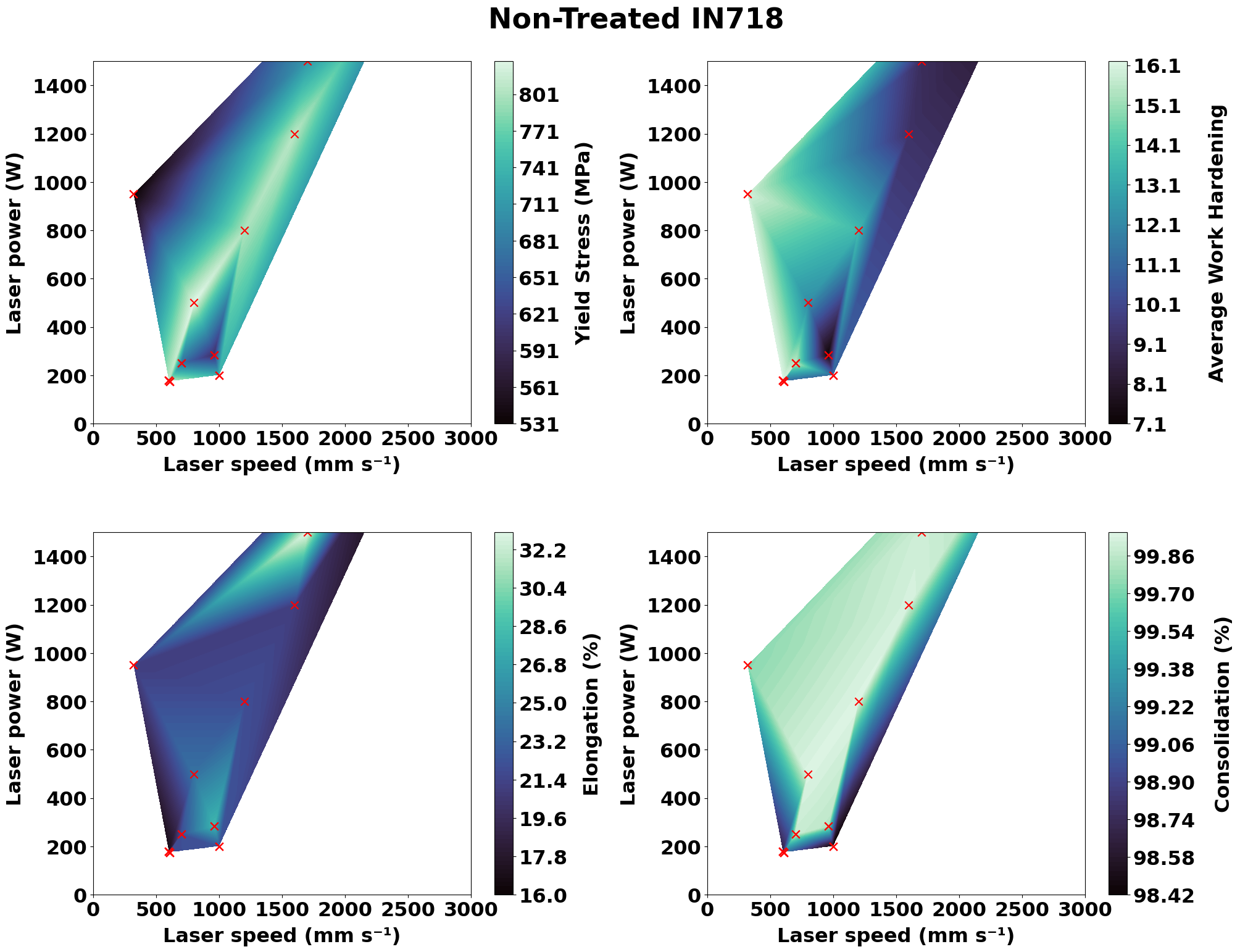}
\caption{Non-treated IN718 $Hmap_1$, individual processing window maps for yield strength, average work-hardening, elongation and consolidation.}
\label{fig:In718 NT - All Properties}
\end{figure}

\begin{figure}[h!]
\centering
\includegraphics[width = 0.75\hsize]{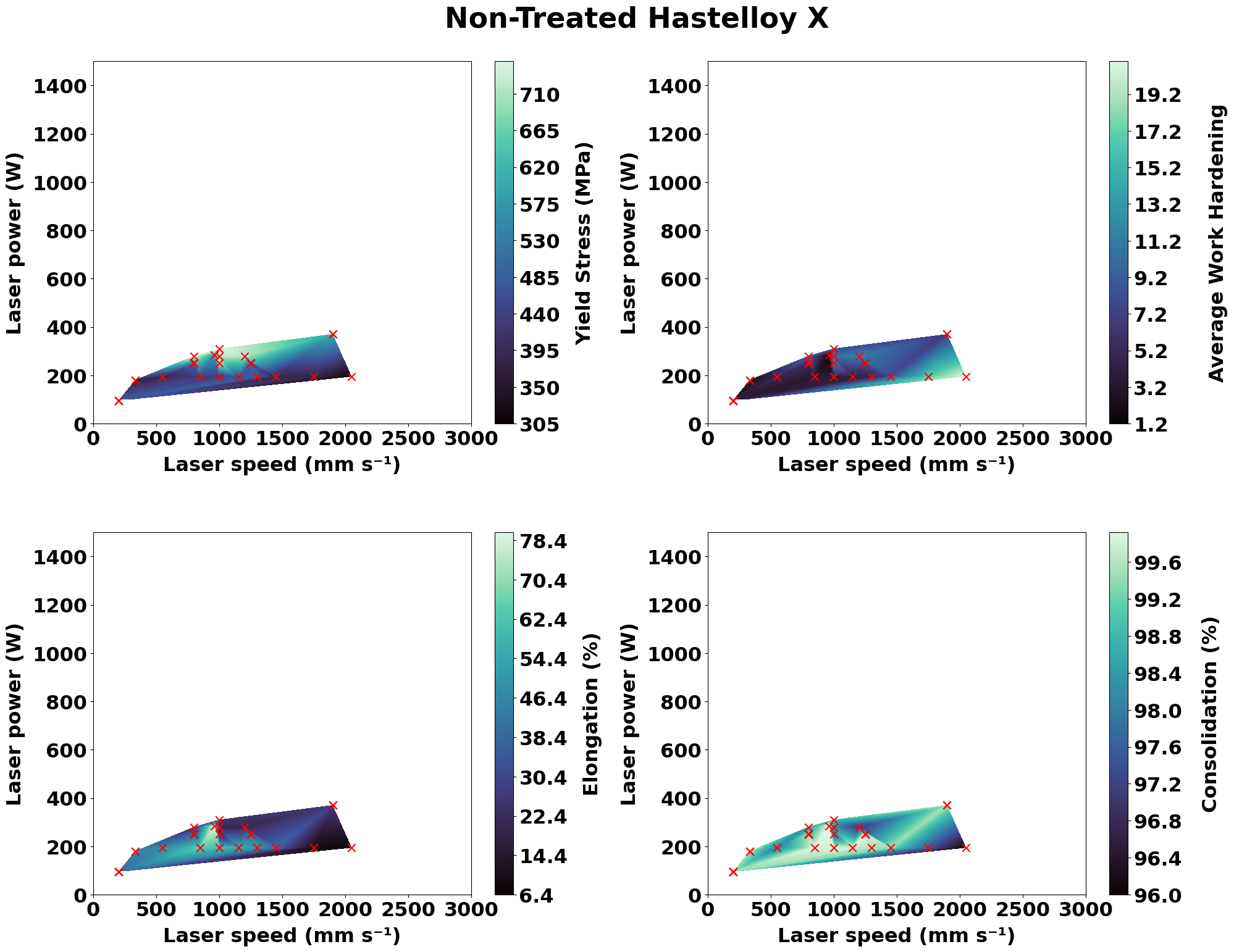}
\caption{Non-treated Hastelloy X $Hmap_1$, individual processing window maps for yield strength, average work-hardening, elongation and consolidation.}
\label{fig:Hastelloy NT - All Properties}
\end{figure}

\clearpage

\bibliographystyle{MSP}
\bibliography{template.bbl}

\end{document}